\documentclass[aps,twocolumn,floats,superscriptaddress,prd,nofootinbib]{revtex4} %
\usepackage[dvips]{graphicx} 
\usepackage{epsfig,amsmath}
\usepackage{amssymb}
\usepackage{bm}
\usepackage{xstring}
\usepackage{rotate}
\usepackage{color}
\usepackage[ps2pdf=true]{hyperref}

\newcommand{\be}{\begin{equation}}
\newcommand{\ee}{\end{equation}}
\newcommand{\bea}{\begin{eqnarray}}
\newcommand{\eea}{\end{eqnarray}}
       
\newcommand{\refeq}[1]{Eq.~(\ref{eq:#1})}

\newcommand{\reffig}[1]{Fig.~\ref{fig:#1}}          
\newcommand{\reftab}[1]{Tab.~\ref{tab:#1}}

\newcommand{\refsec}[1]{\S~\ref{sec:#1}}          
\newcommand{\refapp}[1]{\S~\ref{app:#1}}          
\newcommand{\EpiG}{8\pi G} 
\newcommand{\FpiG}{4\pi G} 
\newcommand{\dM}{\delta M} 

\renewcommand{\v}[1]{\mathbf{#1}}

\newcommand{\ph}{\varphi}

\newcommand{\vn}{\v{\nabla}}
\newcommand{\vx}{\v{x}}

\renewcommand{\d}{\delta}
\newcommand{\D}{\Delta}

\newcommand{\rhob}{\overline{\rho}}

\newcommand{\iMpch}{\,h/{\rm Mpc}}

\newcommand{\Msunh}{\,M_{\odot}/h}

\newcommand{\Om}{\Omega_m}

\newcommand{\OL}{\Omega_\Lambda}
\newcommand{\Orc}{\Omega_{\rm rc}}
\renewcommand{\L}{\Lambda}

\newcommand{\Rvir}{R_{\rm vir}}
\newcommand{\Mvir}{M_{\rm vir}}

\newcommand{\Dvir}{\D_{\rm vir}}
\newcommand{\avir}{a_{\rm vir}}
\newcommand{\cvir}{c_{\rm vir}}

\newcommand{\Rta}{R_{\rm ta}}

\newcommand{\blin}{b_{\rm lin}}

\begin{document}

\title{Spherical Collapse and the Halo Model in Braneworld Gravity}

\author{Fabian Schmidt}
\affiliation{Theoretical Astrophysics, California Institute of Technology M/C 350-17,
Pasadena, California  91125-0001, USA}
\author{Wayne Hu}
\affiliation{Department of Astronomy \& Astrophysics, Kavli Institute for Cosmological Physics, and Enrico Fermi Institute, The University of
Chicago, Chicago, IL 60637-1433}
\author{Marcos Lima}
\affiliation{Department of Physics \& Astronomy, University of Pennsylvania, Philadelphia PA 19104}

\begin{abstract}
We present a detailed study of the collapse of a spherical perturbation
in DGP braneworld gravity for the purpose of modeling simulation results for the halo mass
function, bias and matter power spectrum.  The presence of evolving modifications to the
gravitational force in form of the scalar brane-bending mode lead to
qualitative differences to the collapse in ordinary gravity.  In particular,
differences in the energetics of the collapse necessitate 
a new, generalized method for defining the virial radius
which does not rely on strict energy conservation.  These differences and 
techniques apply to smooth dark energy models with $w\neq -1$ as well.
We also discuss the impact of the exterior of the perturbation on collapse
quantities due to the lack of a Birkhoff theorem in DGP.
The resulting predictions for the mass function, halo bias and power
spectrum are in good overall agreement with DGP N-body simulations
on both the self-accelerating and normal branch.
  In particular, the impact of the Vainshtein mechanism as measured
in the full simulations is matched well.  The model and techniques introduced 
here can serve as practical tools for placing consistent constraints on 
braneworld models using observations of large scale structure.
\end{abstract}


\date{\today}

\maketitle

\section{Introduction}
\label{sec:intro}

Modified gravity models have attracted a great deal of interest
recently as an alternative explanation of the observed accelerated
expansion of the Universe \cite{RiessEtal,KowalskiEtal,WMAP5,Giannantonio,Pietrobon}.  In order for this scenario to work,
gravity must be significantly modified from General Relativity (GR)
on cosmological scales, but has to reduce to GR locally
in order to satisfy stringent Solar System constraints at a few AU.
 Thus, a working modified gravity model has to include a non-linear mechanism
to restore GR in high-density environments, which can have a noticeable
impact on the formation of large-scale structure on intermediate scales of a few to tens of 
Mpc \cite{LueEtal04,HPMpaperII,HPMhalopaper,DGPMpaper}.

One popular modified gravity scenario is the DGP braneworld model \cite{DGP1}.
  Here, a four-dimensional Friedmann-Robertson-Walker universe is imbedded
as a brane in five-dimensional Minkowski space.  In this model, gravity is 
5-dimensional on the largest scales, and becomes four-dimensional at scales 
below the crossover scale $r_c$, a fundamental parameter of the model.  The modification
of the 
Friedmann equation depends on the choice of embedding for the brane \cite{Deffayet01}
which yields two branches of the model: the
\emph{self-accelerating} branch, on which accelerated
expansion occurs at late times without any cosmological constant or
dark energy, and the \emph{normal} branch where no such acceleration
occurs, and a brane tension or other form of stress-energy with negative
pressure has to be added on the brane.  

On scales much smaller than the horizon and crossover scales,
 DGP gravity can be described
by an effective scalar-tensor theory \cite{NicRat,KoyamaMaartens}, where the additional scalar degree
of freedom, the brane-bending mode $\ph$ is associated with displacements 
of the brane from its background position.  
The brane-bending mode yields an
additional gravitational force which influences dynamics of non-relativistic
particles.  On the normal branch of DGP this force is attractive, 
while it is repulsive in the self-accelerating branch.
Non-linear interactions of $\ph$, via the so-called Vainshtein mechanism, 
ensure that it has tiny effects within the Solar System.  
For values of $r_c$ of order the current horizon,
these interactions become important as soon as the density contrast 
becomes of order unity.  Hence, it is crucial to consistently follow
the full brane-bending mode interactions in a cosmological simulation,
as has recently been done \cite{DGPMpaper,ScII}.  
While we study the behavior of $\ph$ in specific DGP models, the form
of the non-linear interactions is expected to be generic to braneworld
models with large extra dimensions \cite{deGrav,galileon}.

Given the considerable
computational expense of these simulations, it is worthwhile to develop
a model which captures the main modified gravity effects, enabling
forecasts and constraints properly marginalized over cosmological parameters.
In the halo model of large structure \cite{CooraySheth}, which assumes that all matter is in bound dark matter halos,
the abundance and clustering of halos are determined
by the linear power spectrum once characteristic quantities of
the collapse of a spherical perturbation are determined, namely, the linear
collapse threshold and the virial radius or overdensity.

Spherical collapse is well studied for General Relativity with a cosmological
constant (e.g., \cite{Peebles80,LahavEtal91}), and has also been explored for
quintessence-type dark energy \cite{WS98,HB05,BasilakosEtal}.  A first study of spherical
perturbations in the
context of DGP was undertaken in \cite{LueEtal04}.  
%
%
%
We extend previous studies by deriving the full $\ph$ field profile of an 
isolated mass in closed form, and by carefully defining the interior and exterior forces during collapse along with the energetics implied by the profile.  In particular, the potential energy required for the virial theorem and for the
total Newtonian energy differ, with the latter not being strictly conserved
during collapse.\footnote{The ``total energy'' here is defined for a Newtonian cosmology and
its non-conservation does not imply violation of covariant energy-momentum conservation.}
 This lack of a conservation law also applies to
dark energy models with an equation of state $w\neq -1$.  We show how this
problem can be circumvented by properly defining the condition for virial equilibrium based
on forces.

We discuss the general properties and parameterization of DGP models in \refsec{models}, and the spherical
collapse calculation in \refsec{rev}.  The halo
model calculations are outlined in \refsec{HM}, and the results
and comparisons with simulations are given in \refsec{res}.  We conclude
in \refsec{concl}. 
 The Appendix contains derivations of the
$\ph$ profile and other quantities needed in the collapse calculation,
as well as a discussion of the potential energy and virial theorem in
DGP.

\section{DGP Models}
\label{sec:models}

In this section, we discuss the general properties of the
 DGP models considered in this paper. The
first model (\emph{sDGP}) is in the self-accelerating branch of DGP with neither
a cosmological constant nor spatial curvature. 
During matter domination and beyond, the 
modified Friedmann equation in sDGP reads:
\bea
H_{\rm sDGP}(a) = H_0 \left (\sqrt{\Orc} + \sqrt{\Om a^{-3} + \Orc} \right ), \label{eq:HsDGP}
\eea
where
\bea  \Orc \equiv \frac{1}{4H_0^2r_c^2}, \quad \Om \equiv \frac{8\pi G}{3 H_0^2} \bar\rho_{0},
\eea
and $\bar\rho_{0}$ is the average matter density today. This expansion history
is clearly different from $\Lambda$CDM, and corresponds to an effective
dark energy with $w_{\rm eff} \rightarrow -1/2$ in the matter-dominated era
at high redshifts. For comparison, we will also consider an effective
smooth dark energy model (\emph{QCDM}) with the same expansion history as 
sDGP.  Note that this expansion history when combined with the growth of structure
near the horizon is in substantial conflict with data \cite{FangEtal}. 
Moreover, the self-accelerating branch is plagued by
ghost issues \cite{LutyEtal,NicRat,KoyamaReview} when perturbed around the de Sitter limit.  
Despite these problems, sDGP remains an interesting toy model for acceleration from 
modified gravity.  

The second scenario (\emph{nDGP}) is in the normal branch of DGP. In order
to achieve acceleration, it is necessary to add a stress-energy component
with negative pressure on the brane. We adopt the model introduced in
\cite{DGPMpaperII}, where a general dark energy component is added on
the brane but the geometry remains spatially flat. The equation of state of this dark energy is adjusted so that
the expansion history is precisely $\L$CDM:
\be
H_{\rm nDGP}(a) = H_0 \sqrt{\Om a^{-3} + \OL},
\label{eq:HnDGP}
\ee
again during matter domination and beyond.
While $\Om$ quantifies the true matter content in this model, $\OL$ is
to be seen as an effective cosmological constant relevant for the
expansion history only.   This construction allows our nDGP models to evade the
otherwise stringent
expansion history constraints on $r_c$ with a true cosmological constant or brane tension
\cite{LombriserEtal}.  Likewise it provides a class of models where the observable impact
of force modification is cleanly separated from the background geometry.   We consider the two models of \cite{DGPMpaperII},
\emph{nDGP--1} with $r_c=500\:\rm Mpc$ and \emph{nDGP--2} with
$r_c=3000\:\rm Mpc$.

For notational simplicity,
it is convenient to also phrase the sDGP Friedmann equation  in terms of an effective
dark energy component so that in both cases
\be
H^2 =\frac{\EpiG}{3}(\rhob + \rho_{\rm eff}),
\ee
where $\rhob$ is the background matter density and $\rho_{\rm eff}$ is implicitly defined by \refeq{HsDGP} and
\refeq{HnDGP}.
In Tab.~\ref{tab:params}, we summarize the parameter choices for the simulated models \cite{DGPMpaper,DGPMpaperII}; in case of sDGP, they are from the best-fitting flat self-accelerating
model of \cite{FangEtal} to WMAP 5yr, Supernova and $H_0$ data, while for the 
nDGP models, the
expansion history and primordial normalization match those of the best-fitting
$\Lambda$CDM model of \cite{FangEtal}.

On scales much smaller than both the horizon $H^{-1}$, and 
the cross-over scale $r_c$, DGP reduces to an effective scalar-tensor
theory with the brane-bending mode $\ph$ representing the scalar.
Time variation in $\ph$ induced by the nonrelativistic motion of the matter
involve the dynamical time and can be neglected
with respect to spatial derivatives in this regime. 
The $\ph$ field then couples to matter by contributing to the
metric potentials $\Psi$, $\Phi$ defined by the line element
\be
ds^2 = -(1+2\Psi) dt^2 + a^2 (1+2\Phi) dx^2
\ee
as
\bea
\Psi &=& \Psi_N + \frac{1}{2}\ph\label{eq:Psi}\\
\Phi &=& -\Psi_N + \frac{1}{2}\ph,
\eea
where $\Psi_N$ is the Newtonian potential determined via
the usual Poisson equation
\be
{\nabla^2}\Psi_N = 4\pi G\d \rho, \label{eq:PsiN}.\\
\ee
Here and throughout, spatial derivatives are physical, not comoving.

While the motion of massive, non-relativistic particles such as 
cold dark matter is governed by the dynamical potential $\Psi$, the propagation of light
is determined by the lensing potential $(\Psi-\Phi)/2$. This combination
is not affected by $\ph$ due to the conformal invariance of electromagnetism.
Hence, in DGP lensing mass is equal to the ``actual'' mass, while the dynamical
mass differs unless $|\ph/\Psi_{N}| \ll 1$.

In the sub-horizon, quasi-static regime, the 
equation for the brane-bending mode can be written as (e.g.,~\cite{KoyamaSilva})
\be
\nabla^2 \ph + \frac{r_c^2}{3\beta} [ (\nabla^2\ph)^2
- (\nabla_i\nabla_j\ph)(\nabla^i\nabla^j\ph) ] = \frac{\EpiG}{3\beta} \delta\rho ,
\label{eq:phiQS}
\ee
where the function $\beta(a)$ is given by
\be
\beta(a) = 1 \pm 2 H(a)\, r_c \left ( 1 + \frac{\dot H(a)}{3 H^2(a)} \right ).
\label{eq:beta}
\ee
Here, the positive sign holds for the normal branch of DGP, while the negative
sign holds for the self-accelerating branch.
%
 Note that the sign of $\beta$ determines whether
the force mediated by the brane-bending mode is attractive ($\beta >0$, nDGP) or repulsive
($\beta < 0$, sDGP).

\begin{table}[b!]
\caption{Parameters of the simulated DGP cosmologies.\label{tab:params}} 
\begin{center}
  \leavevmode
  \begin{tabular}{l||l|l||l|l|l}
\hline
  & QCDM & sDGP & $\Lambda$CDM & nDGP--1 & nDGP--2 \\
\hline
$\Om$ & \  0.258 & \  0.258 & \  0.259 & \  0.259 & \  0.259 \\
$\OL$ (eff.) & \ 0 & \ 0 & \ 0.741 & \ 0.741 & \ 0.741 \\
$r_c$~[Mpc]& \ $\infty$ & \ 6118 & \  $\infty$ & \  500 & \  3000 \\
$\Orc$ & \ 0 & \  0.138 & \  0 & \  17.5  & \  0.487 \\
$H_0$~[km/s/Mpc] & \  66.0 & \  66.0 & \  71.6 & \  71.6 & \  71.6 \\
\hline
$100\,\Omega_b\,h^2$ & \multicolumn{2}{|c||}{2.37} & \multicolumn{3}{c}{2.26}\\
$\Omega_c\, h^2$ & \multicolumn{2}{|c||}{0.089} & \multicolumn{3}{c}{0.110}\\
$\tau$ & \multicolumn{2}{|c||}{0.0954} & \multicolumn{3}{c}{0.0825}\\
$n_s$ & \multicolumn{2}{|c||}{0.998} & \multicolumn{3}{c}{0.959}\\
$A_s(0.05\,{\rm Mpc}^{-1})$ & \multicolumn{2}{|c||}{$2.016\: 10^{-9}$} & \multicolumn{3}{c}{$2.107\: 10^{-9}$}\\
\hline\hline
$\sigma_8(\Lambda\rm CDM)$\footnote{Linear power spectrum normalization today
of a $\Lambda$CDM model with the same primordial normalization.}
  & \multicolumn{2}{|c||}{0.6566} & \multicolumn{3}{c}{0.7892}\\
\hline
\end{tabular}
\end{center}
\end{table}

\section{Top-hat Collapse in DGP}
\label{sec:rev}

In this section, we review the dynamics of the collapse of a top-hat density 
perturbations in the DGP case. We follow \cite{LueEtal04, HPMhalopaper} 
but pay special attention to the $\ph$ profile as well as subleties in the
potential energy and virial condition
which are specific to
the Vainshtein mechanism (see Appendix for details).

We assume an initial top-hat density profile of the form
\be
\rho(r) - \rhob = \left\{
\begin{array}{rl}
\d\rho, & r \leq R,\\
0, & r > R.
\end{array} \right .
\label{eq:drho}
\ee
We show in the Appendix that the top-hat profile remains top-hat during
the collapse despite sweeping out an underdensity outside of $R$.  We further show that
 forces inside
of $R$ depend on the enclosed mass perturbation and so we can ignore the impact of
any compensating underdensity on the dynamics of collapse.
Note that this is unique to a top-hat density and does not hold for other
density profiles, in which case the collapse will not be self-similar anymore.

Given $\rho$ and $R$, there are two important mass parameters: the total mass
$M =4\pi \rho R^3 /3 $ and the mass perturbation
$\delta M = 4\pi  \d\rho R^3/3$.  
The first is conserved during collapse, while 
the second is the source of the $\ph$ field and gravitational
potential, and is therefore useful in expressions involving the field profile.
With the definition $\delta = \d\rho/\rhob$, the two masses are related by 
\begin{equation}
\delta M ={ \delta \over 1+\delta} M ,
\end{equation} 
such that they coincide at high overdensity.

\subsection{Collapse Dynamics}

Given a metric specified by $\Psi$, $\Phi$,
conservation of energy-momentum  is unchanged in DGP and leads to the same equation of motion for the density perturbation as in ordinary gravity.  On scales much smaller
than the horizon
\be
\ddot\d - \frac{4}{3}\frac{\dot\d^2}{1+\d}+2 H\:\dot\d = (1+\d)
\nabla^2 \Psi,
\label{eq:collapseeqn}
\ee
where $H=\dot a/a$ denotes the Hubble rate, and dots denote derivatives
with respect to time.
The modification of gravity enters through the dynamical potential
$\Psi$: in general relativity (GR), $\Psi=\Psi_N$ [\refeq{PsiN}], while in DGP
$\Psi$ receives an additional contribution from the brane-bending mode $\ph$
following \refeq{Psi}.

The full profile of $\ph$ around a top-hat perturbation is derived
in the Appendix. The key result is that in the {\it interior}
of the top-hat, $\nabla^2\ph$ is constant like $\nabla^2\Psi_N$.
Hence, a pure top-hat will stay top-hat, so that the 
\refeq{collapseeqn} can be considered as an ordinary differential equation involving a spatially constant $\delta$. Note that as shown in 
\refapp{profile} this is unique to a top-hat
and will be violated as soon as more general spherically symmetric profiles are considered.
Moreover the implied scaling with the local matter density
 is not true for the exterior of the top-hat (cf.~\cite{KW}).
We discuss this issue further in the Appendix.

While $\nabla^2\ph$ is spatially constant for $r \le R$, it has a nontrivial dependence
on $\d\rho$ (see Appendix). This dependence can be cast in terms
of an effective gravitational constant:
\bea
\nabla^2\ph &=& 8\pi \D G_{\rm DGP}(R/R_*) \d\rho,\label{eq:delphi}\\
\D G_{\rm DGP}(x) &=&  \frac{2}{3\beta} \frac{\sqrt{1+x^{-3}}-1}{x^{-3}} G,
\label{eq:Geff}
\eea
where the Vainshtein radius \begin{equation} 
R_* =\left(\frac{16  G \d M r_c^2}{9\beta^2}\right)^{1/3}.
\label{eq:rV}
\end{equation}
If $R \gg R_*$, $\D G_{\rm DGP} = G/(3\beta)$ and
$\ph$ is simply proportional to the Newtonian potential, $\Psi_N(r) = G\d M/r$.
We call this the linearized limit as it applies to the $\delta M \ll M$ limit.

In the opposite
limit relevant for $\d\rho/\rhob \gg 1$, $\D G_{\rm DGP} \propto (R/R_*)^{3/2} \ll 1$. This
is the Vainshtein limit and here $\ph \sim G\d M/R_*$ is to leading order constant 
throughout the perturbation (\refapp{prof}).

We can use mass conservation
\be
M = \frac{4\pi}{3}R^3\rhob (1+\d)
\ee
to rewrite the top-hat equation of motion in \refeq{collapseeqn} as
\bea
\frac{\ddot R}{R} &=& H^2 + \dot H - \frac{1}{3}\nabla^2 \Psi\nonumber\\
&=& -\frac{4\pi G}{3}\left [ \rhob + (1+3 w_{\rm eff}) \rho_{\rm eff} \right ]\nonumber\\
& & - \frac{4\pi G_{\rm DGP}(R/R_*)}{3}\d\rho,
\label{eq:collapse-r}
\eea
where $G_{\rm DGP}\equiv G + \D G_{\rm DGP}$.
In the second line, the effective equation of state is
defined to be  $-3(1+w_{\rm eff}) \equiv d\ln\rho_{\rm eff}/d\ln a$. These
terms account for the effect of the background expansion.

The terms involving the expansion history can be seen as coming from
an effective potential obtained by expanding the Friedmann-Robertson-Walker
metric around the center of the perturbation \cite{LahavEtal91,HB05}:
\begin{equation}
\Psi_{\rm eff} = -{1\over 2}\left( {\ddot a \over a} \right) r^2 = 
\frac{2\pi G}{3}\left ( \rhob + (1+3w_{\rm eff})\rho_{\rm eff}\right ) r^2\,,
\label{eq:Psieff}
\end{equation}
up to a constant that is irrelevant for the dynamics (see \ref{app:profile}).
Note that $\nabla^2 \Psi_{\rm eff}$ is spatially constant and so its effect also preserves
the top-hat profile. 

\refeq{collapse-r} can then be written in compact form as 
\be
\frac{\ddot R}{R} = -\frac{1}{3} ( \nabla^2\Psi_{\rm eff} + \nabla^2\Psi ).
\label{eq:collapse}
\ee
Note that the pieces involving the matter density combine as
\be
\nabla^2 (\Psi_{\rm eff}+\Psi_N) = \FpiG [ \rho + (1+3w_{\rm eff})\rho_{\rm eff} ],
\ee
so as to reflect the total matter density $\rho$ inside the top-hat as one would expect
from Newtonian mechanics (see Appendix for further discussion).

\subsection{Collapse calculation}
\label{sec:collapse}

We numerically solve the spherical collapse equation~(\ref{eq:collapse-r})  following \cite{HPMhalopaper}.
Specifically, we start at an initial scale factor $a_i=10^{-5}$,
using $\ln a$ as a time variable,
and replacing $R$ with $y$ defined as
\be
y \equiv \frac{R}{R_i} - \frac{a}{a_i},
\ee
where $R_i$ is the initial radius of the perturbation.
Hence, we start with $y=0$ and $y'=-\delta_{i}/3$ as given by linear theory in the
matter dominated epoch in terms of the initial density fluctuation 
$\delta_{i}$.  Here, we have set $\D G_{\rm DGP}=0$ at $a_i$, since the
effects of force modifications in DGP are negligible at such an early time.

With these initial conditions, we can solve \refeq{collapse-r} as
\bea
y'' & =& -\frac{H'}{H} y' + \left(1 + \frac{H'}{H}\right ) y\nonumber\\
& & - \frac{\Om H_0^2 a^{-3}}{2 H^2(a)} {G_{\rm DGP}(R/R_*) \over G} \left(y+\frac{a}{a_i}\right)  \, \d .
\label{eq:collapsey}
\eea
The overdensity relative to the background $\d$ is given by
\be
\d(y,a) = (1+\d_i)\:\left(\frac{a_i}{a} y + 1\right )^{-3} -1.
\ee

\begin{figure}[t!]
\centering
\includegraphics[width=0.48\textwidth]{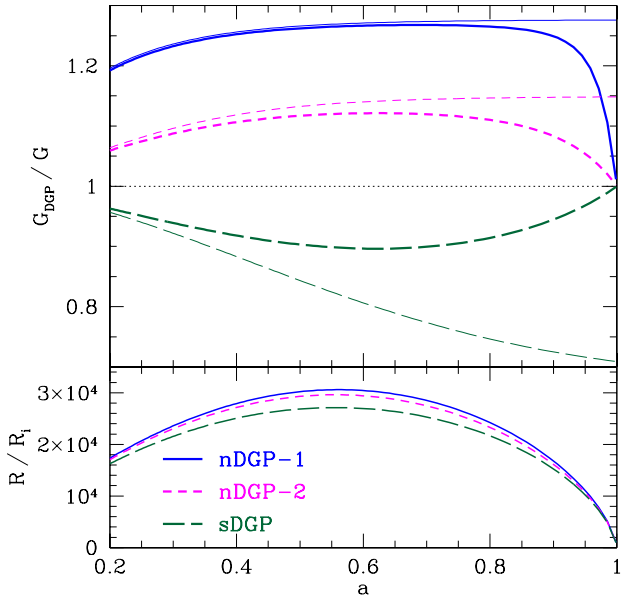}
\caption{Evolution of gravitational force $G_{\rm DGP}/G$ as a function of $a$
for perturbations collapsing at $a_0=1$ in the different DGP models
(top panel).  In each case, the thin lines show the linearized force
modification \refeq{Glin} whereas the thick lines show the nonlinear case. 
We also show the evolution of the scaled radius
$R/R_i$ of the perturbation in the bottom panel.
\label{fig:Gcollapse}}
\end{figure}

\reffig{Gcollapse} shows the result of solving this equation  in the different
models (bottom panel).   We adjust $\d_i$  so that collapse, 
 where $R=0$ or $y = -a/a_i$, occurs at $a=1$. 
 Turnaround, where $dR/d\ln a = 0$ or $y' = -a/a_i$, occurs at 
$a=0.54-0.56$ for these models.

\reffig{Gcollapse} also shows
the evolution of the gravitational force strength
$G_{\rm DGP}/G$.
Since collapse is defined by $\d \rightarrow \infty$ at $a\rightarrow a_0$, 
the perturbation eventually becomes much smaller than its Vainshtein
radius in all models, so that $G_{\rm DGP}\rightarrow G$.   Thus, 
the evolution of forces between turnaround and collapse is significant.
This evolution
raises the issue of the conservation of total energy of the perturbation
during collapse.
We will return to this question in the Appendix.

We then extrapolate the initial overdensity $\d_i$ to $a_0=1$ using
the {\it linear} growth equation, obtained from linearizing 
\refeq{collapseeqn}:
\be
\ddot\d + 2H\:\dot\d = 4\pi G_{\rm lin} \rhob\:\d,
\label{eq:collapse-lin}
\ee
where
\be
G_{\rm lin}(a) = \left [ 1 + \frac{1}{3\beta(a)} \right] G
\label{eq:Glin}
\ee
is the linearized value of $G_{\rm DGP}$ for $R/R_* \gg 1$ in 
\refeq{collapsey}. The resulting overdensity is the linearly extrapolated 
collapse overdensity, which we call $\d_c$. 

To expose the impact of the Vainshtein mechanism, we will also consider 
{\it linearized DGP} collapse
(note that this it \emph{not} linearized collapse) as the limit
of no non-linear $\ph$ interactions. 
In this case, we replace $G_{\rm DGP}$ with $G_{\rm lin}$ in 
\refeq{collapsey}.
Note that the Vainshtein mechanism is strongest for a spherically symmetric collapse
and absent for a planar collapse and so these two cases should encompass the
range of possibilities in the cosmological context \cite{DGPMpaper,ScI}.

\begin{figure}[t!]
\centering
\includegraphics[width=0.48\textwidth]{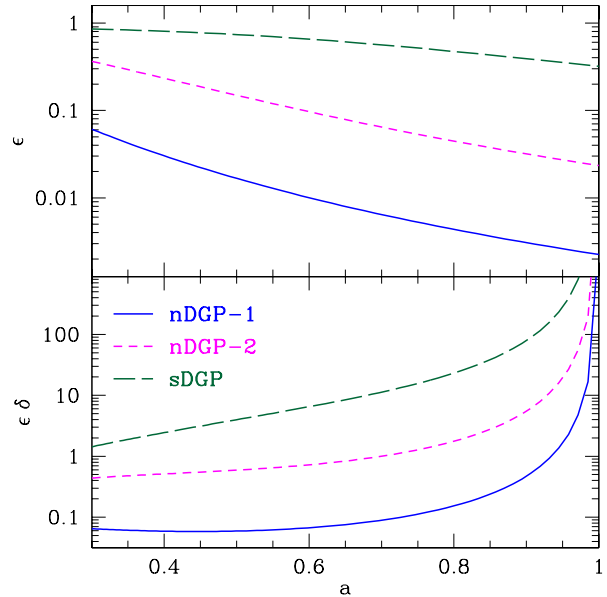}
\caption{\textit{Top panel:} The non-linearity parameter $\epsilon$.
The Vainshtein mechanism operates once
$\epsilon\,\d \gtrsim 1$.
\textit{Bottom panel:} $\epsilon\,\d$ for a top-hat perturbation that
collapes at $a_0=1$ in each model. 
\label{fig:beta}}
\end{figure}

For a top-hat, the Vainshtein radius in units of $R$ is given by
$R/R_* = (\epsilon \delta)^{-1/3}$, where $\epsilon$ is defined as
\be
\epsilon = \frac{8}{9\beta^2(a)}(H_0 r_c)^2\: \Om\:a^{-3},
\label{eq:eps}
\ee
such that $1/\epsilon$ represents the density threshold at which the top-hat perturbation
crosses into its own Vainshtein radius.
\reffig{beta} shows $\epsilon$ and $\epsilon\delta$
as a function of $a$ for a range between turnaround $(a \sim 0.5)$ and collapse
$(a = 1)$
for the simulated models of Tab.~\ref{tab:params}.   
In the nDGP models
this threshold increases substantially toward the present and the Vainshtein suppression
does not saturate  until quite late in the collapse.  In fact in the nDGP-1 model, we shall
see that the perturbation virializes before the Vainshtein mechanism can operate.

\subsection{Virial Radius and Overdensity}
\label{sec:virial}

Spherical top-hat collapse formally predicts a collapse to a singularity at $R \rightarrow 0$.  
In reality we expect that processes such as
violent relaxation will eventually establish virial equilibrium.  
More specifically, the virial theorem relates the kinetic energy of the body
\be
T = \int d^3x\:\frac{1}{2}\rho v^2 = \frac{3}{10} M\dot R^2,
\ee
where the last equality holds for a top-hat, to
the trace of the potential energy tensor
\begin{equation}
2T + W = 0,
\label{eq:virialtheorem}
\end{equation}
with the definition
\bea
W &\equiv&  -\int d^3x \:\rho_{m} (\vx)\: \vx\cdot\vn\Psi .
\eea
Note that $W$ depends explicitly on forces only and some care must be taken in
relating it to the potential or binding energy of the perturbation for contributions from the 
brane bending mode $\ph$ (see Appendix \ref{app:virialeq}). 
We show there that $W$ can be broken up as a sum of three contributions to
the dynamics,
\begin{eqnarray}
W &=& -{3 \over 5} {GM^{2}\over R} - {4\pi G \over 5} (1+3 w_{\rm eff})\rho_{\rm eff} M R^2 \nonumber\\
 && -{3 \over 5} {\Delta G_{\rm DGP} M \delta M \over R} .
\label{eq:Wtot}
\end{eqnarray}

We determine the virial radius $\Rvir$ of the perturbation as the radius during collapse (after turnaround)
at which \refeq{virialtheorem} is satisfied.
In the literature conservation of total energy is often used to set this condition.  One can easily show that if energy conservation holds strictly, our
determination of $\Rvir$ agrees with the usual definition. However,
in the presence of an evolving $\rho_{\rm eff}$ and $\D G_{\rm DGP}$, energy is no longer
strictly conserved over a Hubble time.   Note that this problem occurs in dark energy
models as well but is usually ignored under the assumption that $w_{\rm eff} \approx -1$.
However, this procedure is not justified when considering modifications due to a finite $1+w_{\rm eff}$.
Differences
between our definition of $\Rvir$ and the one relying on energy conservation
are of the same order as the differences induced by $1+w_{\rm eff}$.
  We discuss these issues further in 
Appendix \ref{app:PEusage}-\ref{app:Econservation}.

Once $\Rvir$ is calculated, we 
have the density of the perturbation at this radius, 
$\rho_{\rm vir}=\rhob(\avir)[1+\d(\Rvir)]$, where $\avir$ is the scale
factor at which the perturbation reaches $\Rvir$ during collapse.  For collapse at
$a_0=1$, $\avir \approx 0.91-0.93$ for the simulated models.
One then assumes that the perturbation maintains this ``virial density'' while the
background continues to decrease. 
The final virial density with respect to the background,
$\Dvir$ at $a_0$ is then given by referring
this density to the background matter density at $a_0$:
\be
\Dvir = [1+\d(\Rvir)] \left ( \frac{a_0}{\avir} \right )^3.
\ee
\reftab{sc_param} shows the resulting spherical collapse parameters,
$\d_c$ and $\Dvir$, for the different models and gravitational modifications:
unmodified ({\it GR collapse}), valid for GR with smooth dark energy;
the expression \refeq{Geff} 
({\it DGP collapse}); and \refeq{Glin} ({\it linearized DGP collapse}).

\begin{table}[b!]
\caption{Spherical collapse parameters for the cosmologies defined in
\reftab{params} for collapse at $a_0=1$.\label{tab:sc_param}} 
\begin{center}
  \leavevmode
  \begin{tabular}{l|l|l|l|l}
\hline
& \,  Collapse type/Model: \   & sDGP & nDGP-1 & nDGP-2 \\
\hline\hline
$\d_c$ & GR    & 1.662 & 1.674 & 1.674 \\
 & DGP         & 1.627 & 1.687 & 1.688 \\
 & DGP lin.    & 1.676 & 1.678 & 1.672 \\
\hline\hline
$\Dvir$ & GR   & 399.9 & 372.3 & 372.3 \\
 & DGP         & 467.1 & 300.4 & 322.8 \\
 & DGP lin.    & 436.4 & 311.7 & 339.1 \\
\hline
\end{tabular}
\end{center}
\end{table}


\section{Halo Model Predictions}
\label{sec:HM}

We now briefly describe how we move from spherical collapse predictions
of the linear collapse threshold $\d_c$ and the virial overdensity
$\Dvir$ summarized in \reftab{sc_param}
to predictions of the halo mass function, bias, and non-linear
power spectrum. For further details, see \cite{HPMhalopaper}.

In the Press-Schechter approach, one assumes that all regions with $\d>\d_c$
in the linearly extrapolated initial density field collapse to form bound
structures (halos). The fraction of mass within halos at a given mass is
then determined by the variance of the linear density field smoothed at 
that scale. Here, we adopt
the Sheth-Tormen (ST) prescription \cite{SheTor99}
for the halo mass function predictions, which enables a direct
use of our spherical collapse results. Also, we previously found a good match to the 
ST mass function and bias in our $\Lambda$CDM simulations \cite{HPMhalopaper}.

The ST description 
 for the comoving number density of halos per logarithmic interval in the virial mass $\Mvir$ is given by
\begin{align}
n_{\ln \Mvir} \equiv
\frac{d n}{d\ln \Mvir} &= {\rhob \over \Mvir} f(\nu) {d\nu \over d\ln \Mvir}\,, 
         \label{eq:massfn}
\end{align}
where the peak threshold $\nu = \delta_c/\sigma(\Mvir)$ and 
\begin{eqnarray}
\nu f(\nu) = A\sqrt{{2 \over \pi} a\nu^2 } [1+(a\nu^2)^{-p}] \exp[-a\nu^2/2]\,.
\label{eq:nufnu}
\end{eqnarray}
Here, the virial mass is defined as the mass enclosed at 
the virial radius $\Rvir$.
$\sigma(M)$ is the variance of the linear density field convolved with a top hat of radius $R$
that encloses $M=4\pi R^3 \rhob/3$ at the background density
\begin{eqnarray}
\sigma^2(R) = \int \frac{d^3k}{(2\pi)^3} |\tilde{W}(kR)|^2 P_{\rm L}(k)\,,
\label{eq:sigmaR}
\end{eqnarray}
where $P_{\rm L}(k)$ is the linear power spectrum and $\tilde W$ is the Fourier transform
of the top hat window.  The normalization constant $A$ in \refeq{nufnu} is chosen 
such that $\int d\nu f(\nu)=1$. We adopt the standard parameter values of 
$p=0.3$ and $a=0.75$ throughout.

The linear bias corresponding to the ST mass function, obtained
in the peak-background split, is given by \cite{SheTor99}
\begin{eqnarray}
b_{\rm lin}(\Mvir) & \equiv & b(k=0,\Mvir) \nonumber\\
&=&  1 + {a \nu^2 -1 \over \delta_c}
         + { 2 p \over \delta_c [ 1 + (a \nu^2)^p]}\,.
\label{eq:bias}
\end{eqnarray}

By assuming a specific form of halo density profiles, we can rescale mass 
definitions from the virial mass $\Mvir$ to $M_{200}$, the
mass definition used in the simulation measurements,
as outlined in \cite{HuKravtsov} (again, all overdensities are referred to
the background \emph{matter} density).  
We use this approach 
to compare the scaling relation predictions to the simulations in \S \ref{sec:res}.
For the halo profiles, we take an NFW form \cite{NavFreWhi97},
\begin{equation}
\rho_{\rm NFW}(r) = \frac{\rho_s}{r/r_s (1+ r/r_s)^2},
\label{eq:NFW}
\end{equation}
where $r_s$ is the scale radius of the halo
and the normalization $\rho_s$ is given
by the virial mass $\Mvir$. We parametrize $r_s$
via the concentration $\cvir\equiv \Rvir/r_s$ given by 
\cite{Buletal01}
\begin{equation}
\cvir(\Mvir,z=0) = 9 \left (\frac{\Mvir}{M_*} \right )^{-0.13},
\label{eq:cvir}
\end{equation}
where $M_*$ is defined via $\sigma(M_*)=\delta_c$. Since generally
$\Rvir,R_{200} \gg r_s$, the precise form of the concentration relation
has a negligible impact on the mass rescaling.
In the following, when no specific overdensity is given we implicitly 
take $M=M_{200}$, e.g. 
\begin{equation}
n_{\ln M} \equiv {d n \over d{\ln M_{200}}} = n_{\ln \Mvir} {d{\ln \Mvir}  \over d{\ln M_{200}}} \,.
\end{equation}
We also consider the non-linear matter power spectrum calculated in the 
halo model approach  (see \cite{CooraySheth} for a review). Since all
matter is assumed to be within bound halos, the matter power spectrum can be
decomposed into  1-halo and 2-halo terms,
\be
P_{\rm mm}(k) = I^2(k) P_{\rm L}(k)  + P^{1{\rm h}}(k),
\label{eq:Pkhalo}
\ee
where
\bea
P^{1{\rm h}}(k) &=& \int d\ln \Mvir\: n_{\ln \Mvir} \frac{\Mvir^2}{\bar{\rho}_{\rm m}^2} |y(k,\Mvir)|^2\,,\quad\\
I(k) &=& 
 \int d\ln \Mvir\; n_{\ln \Mvir} \frac{\Mvir}{\bar{\rho}_{\rm m}} y(k,\Mvir) b_{\rm lin}(\Mvir)
 \nonumber \,.
\eea
Here, $y(k,M)$ is the Fourier transform of an NFW density profile truncated at 
$\Rvir$, and normalized so that $y(k,M)\rightarrow 1$ as $k\rightarrow 0$.
Note that with the ST mass function and bias, $\lim_{k \rightarrow 0} I(k) = 1$.




\section{Results}
\label{sec:res}

We compare our spherical collapse and halo model predictions with
the results of N-body simulations presented in \cite{DGPMpaperII,DGPMpaper}
of the sDGP and 
nDGP+DE models (see \refsec{models}, \reftab{params}).   In addition
to the full simulations which solve the non-linear $\ph$ equation~(\ref{eq:phiQS}), simulations using the linearized $\ph$ equation have been performed through 
\refeq{Glin}.

We always compare observables
measured in the DGP simulations with those of GR simulations with the same
initial conditions and expansion history. In this way, cosmic variance as well as systematic issues 
cancel out to a large extent.

\begin{figure}[t!]
\centering
\includegraphics[width=0.48\textwidth]{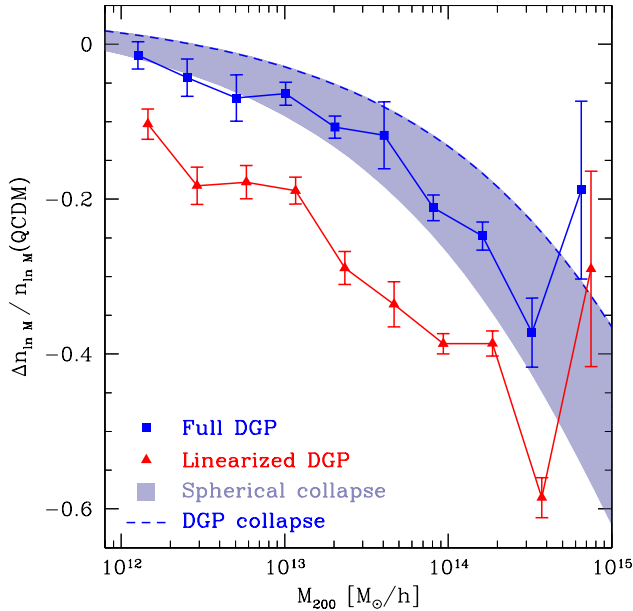}
\caption{Deviation in the halo mass function
at $z=0$
of sDGP from a dark energy model with the same expansion history (QCDM).
The points show measurements in the full and linearized DGP simulations,
the band shows the Sheth-Tormen + spherical collapse prediction range
between full DGP collapse (blue dashed line) and linearized DGP collapse. 
\label{fig:dndm-s}}
\end{figure}
\begin{figure}[t!]
\centering
\includegraphics[width=0.48\textwidth]{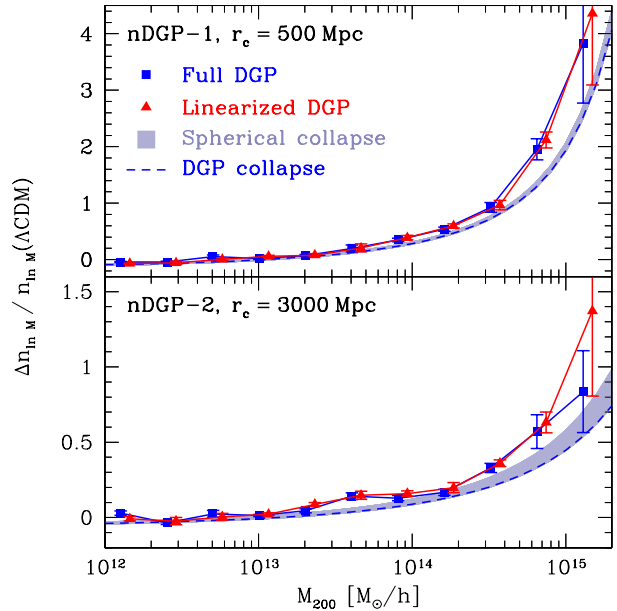}
\caption{Same as \reffig{dndm-s}, for the two normal branch DGP
+ dark energy models nDGP--1 (top) and nDGP--2 (bottom) relative to
$\Lambda$CDM. The linearized
DGP simulation results have been displaced horizontally for clarity.
\label{fig:dndm-n}}
\end{figure}

\subsection{Halo Mass Function}
\label{sec:dndm}

\reffig{dndm-s} shows the deviation of the halo mass function from QCDM
measured in the sDGP and linearized sDGP simulations, and the spherical
collapse predictions. \reffig{dndm-n} shows the corresponding results
for the nDGP+DE models.  The spherical collapse predictions work well in both cases
and in particular match the shape of the deviations and the relative impact of the Vainshtein mechanism.   In both cases, they somewhat underestimate the size of the deviations
at a fixed mass, corresponding roughly to a shift in $\lg M_{200}$ of $\sim 0.3-0.5$.

In the nDGP models, force modifications are stronger (see \reffig{beta}),
leading to larger deviations in the mass function from the corresponding
GR model with the same expansion history. In particular, the abundance of massive halos $M_{200}\gtrsim 10^{14}\Msunh$ is significantly enhanced.
This behavior is due to the exponential sensitivity of the mass function
to $\nu = \d_c/\sigma(M)$ at the high-mass end, and is
similar to what was seen in the large-field $f(R)$
models in \cite{HPMhalopaper}.

Furthermore, the density threshold $\epsilon^{-1}$ for the onset of the 
Vainshtein mechanism is higher in the nDGP models than the sDGP model,
so that the mass function is
less affected by the Vainshtein mechanism in nDGP. This is borne out by both 
simulations and spherical collapse predictions: the relative spread in the 
predictions between the DGP and linearized DGP case shrinks considerably
when going from large to small $r_c$, i.e. from sDGP to nDGP--2 to nDGP--1.

Finally, since the full spherical collapse predictions always slightly underestimate the deviations,
they can be used to place conservative limits on DGP braneworld scenarios 
from measurements
of the halo mass function e.g. from massive clusters \cite{Vikhlinin,RozoEtal09,MantzEtal09}.  Alternatively, the predictions can be recalibrated
based on simulations by introducing a constant shift in $\lg M_{200} \sim 0.3-0.5$.

\subsection{Halo Bias}
\label{sec:bias}
\begin{figure}[t!]
\centering
\includegraphics[width=0.48\textwidth]{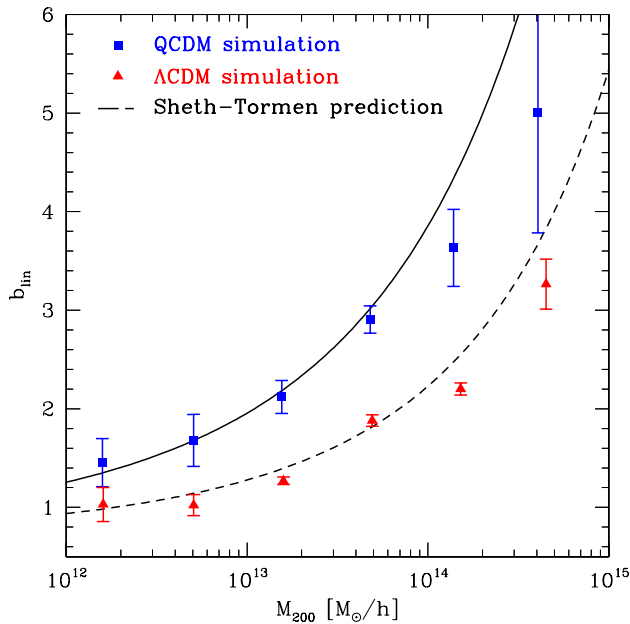}
\caption{Linear bias $\blin$ of dark matter halos measured in the
$\Lambda$CDM and QCDM simulations at $z=0$ (see \refsec{bias}),
and the prediction of the Sheth-Tormen prescription. 
\label{fig:hb-Q}}
\end{figure}
\begin{figure}[t!]
\centering
\includegraphics[width=0.48\textwidth]{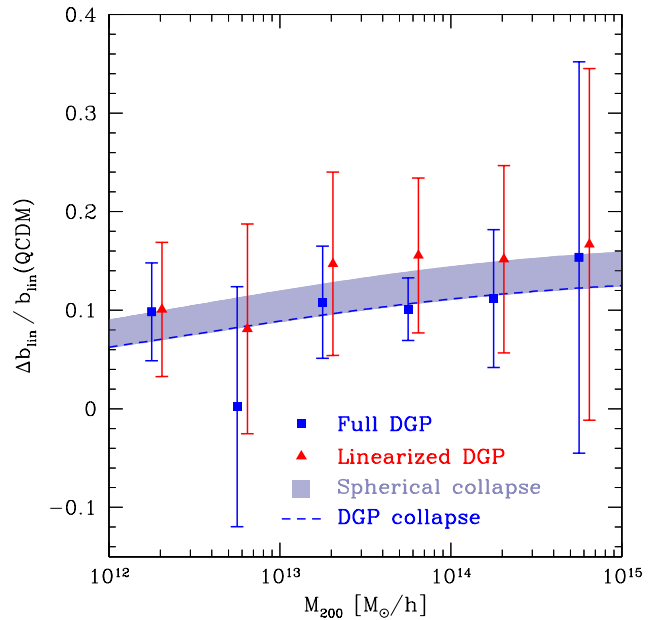}
\caption{Relative deviation of the linear bias $\D\blin/\blin$
in the full and linearized sDGP simulations from that of the 
QCDM simulations as a function of halo mass at $z=0$.  The linearized
simulation points have been displaced horizontally for clarity.
The band shows the Sheth-Tormen + spherical collapse prediction range
between full DGP collapse (blue dashed line) and linearized DGP collapse.
\label{fig:hb-s}}
\end{figure}

This section presents new results on the clustering of halos in the
simulations of \cite{DGPMpaper,DGPMpaperII}.
We extract the {\it linear} halo bias
$b_{\rm lin}(M)$ from our simulations as described in \cite{HPMhalopaper}.
For halos of a given 
logarithmic mass range in a box of size $L_{\rm box}$, we first obtain the 
halo bias $b(k,M)$ by dividing the halo-mass cross spectrum $P_{\rm hm}(k)$ 
by the matter power spectrum for each simulation run.\footnote{Note that the definition of bias adopted will differ from
alternate choices such as $(P_{\rm hh}/P_{\rm mm})^{1/2}$ or $P_{\rm hh}/P_{\rm hm}$
in the non-linear regime where the correlation coefficient between halos
and matter can differ from unity.}
In order to remove trends from the non-linearity of the bias, we then fit 
a linear relation to 
$b(k,M)=b_{\rm lin}(M)+a(M)\: k$ between the minimum $k$ (the fundamental
mode of the box) and $\sim 15 k_{\rm min}$, where 
$b(k,M)$ is the combined measurement from all boxes. 
The same fitting procedure is applied to the run-by-run ratio
of $b_{\rm DGP}(k,M)/b_{\rm GR}(k,M)$.

We then bootstrap over many realizations of the set of simulations, 
performing the fit for every realization. We use the average of the
fit parameter $b_{\rm lin}(M)$ as estimate of the linear bias, and its
spread as an estimate of the error. Note that in case of the nDGP simulations,
we only have 3 runs per box size, so that the error estimate itself has
a large uncertainty. 

We show the linear bias $b_{\rm lin}(M)$ as a function
of halo mass for the QCDM and $\Lambda$CDM simulations themselves in \reffig{hb-Q}.
As the halo mass function deviates significantly
from a pure power-law, especially at the high-mass end, we plot
the bias measurements at the position of the measured average $\lg M_{200}$ of
the halos. The Sheth-Tormen prediction of \refeq{bias}, using the parameters
from \reftab{sc_param} and rescaled from
$\Mvir$ to $M_{200}$ in each case, matches the simulations well for
masses up to $10^{14}\Msunh$. At higher masses it overpredicts the bias
in the simulations, though the deviation is a the level of $1-2\sigma$.
Note that halos are more biased at a given mass in QCDM than in $\L$CDM,
due to the reduced growth and smaller power spectrum amplitude in this model.

\begin{figure}[t!]
\centering
\includegraphics[width=0.48\textwidth]{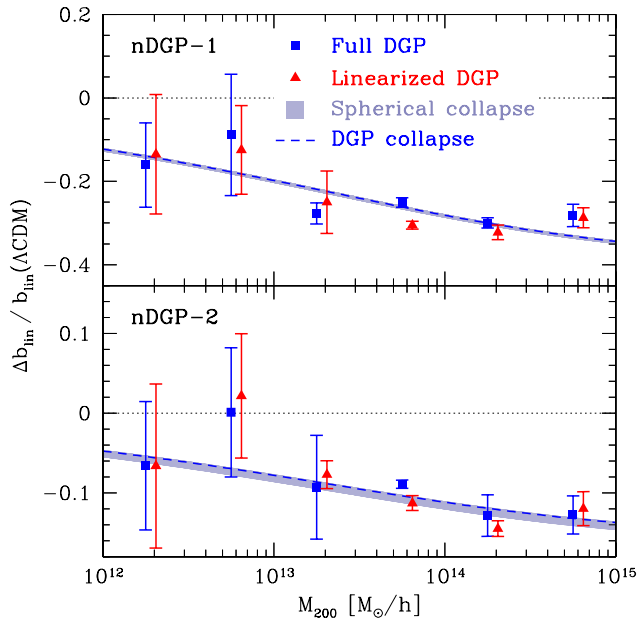}
\caption{Same as \reffig{hb-s}, for the two normal branch DGP
+ dark energy models nDGP--1 (top) and nDGP--2 (bottom), relative to
$\Lambda$CDM.
\label{fig:hb-n}}
\end{figure}

This trend continues in the DGP simulations: in sDGP, gravity is further
weakened by the repulsive brane-bending mode, so that the linear halo bias
at fixed mass is increased by $\sim 10$\% compared to QCDM (\reffig{hb-s}).
There is a hint that halos are slightly higher biased in the
linearized simulations compared to the full simulations, though both are consistent
given the error bars. The spherical collapse prediction matches the
simulation results over the whole mass range. The full DGP collapse
(blue dashed line) marks the lower edge of the shaded band, while the
linearized DGP collapse corresponds to the upper edge, in accordance with
expectations.   The spread of the predictions is similar in magnitude to
the tentative differences between linearized and full simulations.

For the nDGP models, the opposite trend in the bias is seen (\reffig{hb-n}): 
halos are less biased at a given mass. While the full and linearized 
simulation results are consistent in the 
nDGP models, showing no sign of the Vainshtein mechanism affecting the
linear halo bias, they both follow the spherical collapse prediction very
well.  This is in accordance with the good description of the mass function
results in \refsec{dndm}.  Also, given the small spread in the spherical 
collapse predictions, we do not expect to see any differences between full and
linearized simulations for the nDGP models.

\subsection{Non-linear matter power spectrum}
\label{sec:Pk}

We can now assemble the ingredients of the halo model: the mass function,
bias, and profile of halos, to predict the non-linear matter power
spectrum following \refsec{HM}. We assume that the inner parts of the 
density profiles of halos are not affected by DGP, as indicated by
simulations \cite{DGPMpaperII}. More precisely, one can assume that
for a halo of given mass $M_{200}$, the scale radius $r_s$ (\refeq{NFW})
is the same in DGP as in GR. Adopting \refeq{cvir} for the concentration 
$\cvir = \Rvir/r_s$ in GR, we have for the concentration relation
in DGP:
\be
c_{\rm vir,DGP}(M) = \frac{R_\D(M)\big |_{\Dvir({\rm DGP})}}{R_\D(M)\big|_{\Dvir({\rm GR})}}
\cvir(M_{\rm vir,GR}),
\label{eq:cscaling}
\ee
where $R_\D(M) = (3 M / (4\pi \D\rhob))^{1/3}$, and
$M_{\rm vir, GR}$ is the virial mass in GR that corresponds to a virial
mass of $M$ in DGP. We find that for all DGP models considered here,
the concentration $c_{\rm vir,DGP}$ defined by \refeq{cscaling}
is within 3\% of the standard relation
$\cvir(M_{\rm vir,DGP})$, which has a negligible effect on the power
spectrum on the scales probed by the simulations.  Hence, we leave the concentration
relation \refeq{cvir} unchanged in our power spectrum predictions.

\begin{figure}[t!]
\centering
\includegraphics[width=0.48\textwidth]{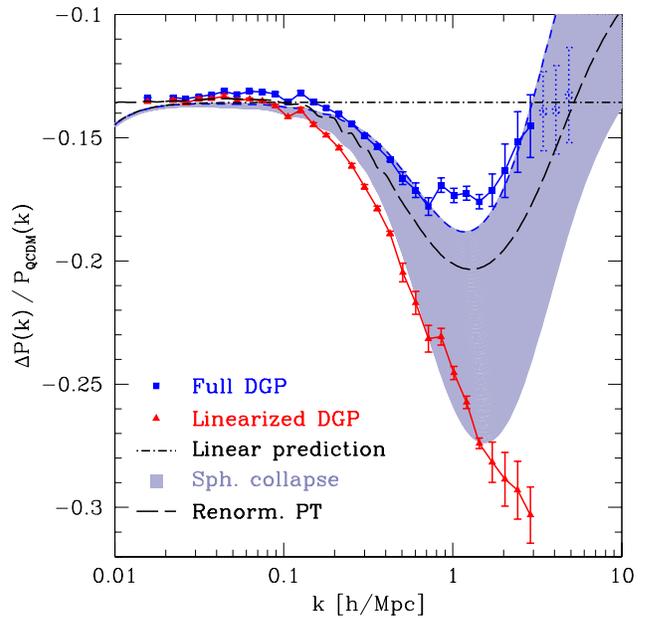}
\caption{Deviation in the matter power
of the sDGP model from QCDM at $z=0$. The points show measurements in the
full and linearized DGP simulations, while the band shows the halo
model prediction based on spherical collapse and the Sheth-Tormen prescription
(between DGP [blue dashed line] and linearized DGP collapse).
The long-dashed line shows the renormalized perturbation theory prediction
from \cite{ScI}.
\label{fig:Pk-s}}
\end{figure}
\begin{figure}[t!]
\centering
\includegraphics[width=0.48\textwidth]{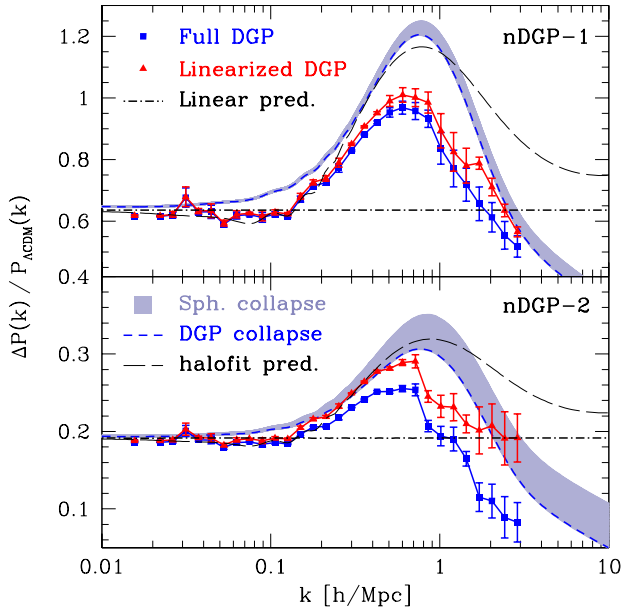}
\caption{Same as \reffig{Pk-s}, for the two normal branch DGP +
dark energy models nDGP--1 (top) and nDGP--2 (bottom) relative to $\Lambda$CDM.
The long-dashed lines show the predictions of {\tt halofit} using the
linear DGP power spectrum.
\label{fig:Pk-n}}
\end{figure}

\reffig{Pk-s} shows that the sDGP simulation results are matched very
well: the DGP collapse prediction (blue dashed line) is close
the to full simulation results, while the linearized DGP collapse
is close to the linearized simulations. The renormalized perturbation
theory prediction of \cite{ScI} which uses a completely different approach
to take into account the non-linear interactions of the brane-bending mode
is also quite close to our DGP collapse calculation.

The match to the power spectrum in the nDGP simulations (\reffig{Pk-n})
is somewhat worse, showing discrepancies of $\sim 30$\% for nDGP--1
and $\sim 10$\% for nDGP--2. In particular, discrepancies can be seen in the
quasi-linear to non-linear transition, $k\sim 0.1-1\iMpch$ for nDGP--1.
The simplified 1-halo/2-halo split in the halo model breaks down on these
scales.  We also show the predicted non-linear power spectra using 
{\tt halofit} \cite{halofit} in combination with the linear DGP
power spectra at $z=0$.  Note that for $k \gtrsim 0.01 \iMpch$, the linear 
DGP power spectrum is identical to that of a $\Lambda$CDM model with
higher linear normalization.  While the match to the simulations is better 
than our spherical collapse predictions at $k \lesssim 0.1\iMpch$, the
deviations grow towards smaller scales so that {\tt halofit} is a worse
fit to the simulations than the spherical collapse model for $k\gtrsim 1 \iMpch$.  
Clearly, it is not trivial to model the simulation results for $\D P(k)/P(k)$ 
to better than $\sim 10$\%.

While the overall magnitude of the power spectrum enhancement
in nDGP is not matched, the shape of $\D P(k)/P(k)$ is matched quite
well.  Furthermore, the effect of the Vainshtein mechanism on the
power spectrum is predicted accurately, as can be seen by comparing
the spread in the spherical collapse predictions to the difference
between full and linearized simulation results.  These findings indicate
that it should be possible to rescale linearized DGP simulation results with the relative Vainshtein
suppression calculated from spherical collapse.

\section{Conclusions}
\label{sec:concl}

By studying the collapse of a spherical perturbation under DGP braneworld
gravity, we have shown how simulation results on the mass
function, halo bias and power spectrum can be understood semi-analytically.
In DGP gravity, force modifications are carried by the scalar brane-bending
mode.  The global properties of the response of the brane-bending mode to matter
control how the Vainshtein mechanism modifies
force and energy conditions during collapse.  These conditions are important for the calculation of virial equilibrium (see the Appendix for detailed
discussions of these results).

In particular, the presence of evolving modifications to the
gravitational force either through the brane bending mode or through the
background expansion violate conservation of Newtonian total energy for
traditional definitions of the potential energy contribution.  This violation applies
to smooth dark energy models with $w\neq -1$ as well.   We introduce a new, general
technique for defining the virial radius
which does not rely on strict energy conservation. 

Under the halo model, these spherical collapse predictions give rise to predictions
for the mass function, halo bias and power
spectrum.  We have shown that these predictions are in good qualitative agreement with DGP N-body simulations
on both the self-accelerating and normal branch.
 In particular, the use of spherical collapse for the mass function always provides slightly
conservative limits on mass function deviations when compared with the simulations.  Hence the semi-analytic techniques introduced here can be used as a practical
tool for extending simulation results for the purpose of studying parameter constraints
on braneworld models from observations of the mass function.

While the absolute power spectrum agreement is not quite as good, these techniques can still be
useful in 
combination with \emph{linearized} DGP simulations. 
Since our spherical collapse predictions appear to capture
accurately the impact of the Vainshtein mechanism on the power spectrum, they could provide an effective way of taking non-linear interactions into account in results obtained
from linearized DGP simulations.
Such simulations are very easy to
implement and an order of magnitude cheaper computationally than the full
DGP simulations. Likewise they are more readily extendable to higher resolution
and can be used to cover a wider range in parameter space.

\acknowledgments

We would like to thank Mark Wyman for a careful reading of the paper
and pointing out typographical errors.

The simulations used in this work have been performed on the Joint 
Fermilab - KICP Supercomputing Cluster, supported by grants from Fermilab,
Kavli Institute for Cosmological Physics, and the University of Chicago. 
F.S. acknowledges support from the Gordon and Betty Moore Foundation
at Caltech.
This work was supported by the Kavli Institute for Cosmological 
Physics at the University of Chicago through grants NSF PHY-0114422 and 
NSF PHY-0551142.  WH was additionally supported  by DOE contract DE-FG02-90ER-40560 and the Lucile and David Packard Foundation.
ML was supported by an NSF-PIRE grant.

\bigskip
\bigskip
\appendix

\section{Top-Hat Overdensity in DGP}
\label{app:tophat}

In this Appendix we present in detail the various aspects of top-hat perturbations in DGP
used in the main text.   We begin by
reviewing the techniques introduced in Ref.~\cite{LueEtal04} for the brane bending
field
but pay special attention to the matching of the various solutions as well as derive a closed form expression for the global field profile.  This global profile
is used to study virial equilibrium (\S \ref{app:virialeq}) and 
potential energy (\S \ref{app:Ebrane}).  We discuss subtleties due to the relationship between the
two and violations of Newtonian energy conservation in \S \ref{app:PEusage} and
their impact on defining the virial overdensity in \S \ref{app:Econservation}.  Finally we
examine the impact of density compensation in the exterior of the top-hat in \S \ref{app:profile} in light of the lack of a Birkhoff theorem in DGP.

\subsection{Spherical Symmetry}

We begin by assuming that the density perturbation $\delta \rho(r)$ is spherically symmetric
but otherwise arbitrary. Given the induced spherical symmetry in the field
solution, 
the field 
equation~(\ref{eq:phiQS}) reduces to
\begin{eqnarray}
&&{1 \over r^{2}}{d \over dr} r^{2}{d \ph \over dr}+
\frac{r_c^2}{3\beta} 
\left[
{4 \over r} {d^2 \ph \over dr^2 } {d\ph \over dr} + 
2 \left( { 1\over r} {d \ph \over dr}\right)^2 \right] \nonumber\\ &&\qquad
= \frac{\EpiG}{3\beta} \delta\rho .
\label{eq:sphericalfield}
\end{eqnarray}
Integration of the field equation (\ref{eq:sphericalfield}) 
 over $r^2 dr$ then yields
\begin{equation}
r^2 {d \ph \over dr} + \frac{2}{3\beta} r_c^2 r ({d \ph \over dr})^2 = \frac{2}{3\beta} Gm(r),
\label{eq:phiSph}
\end{equation}
with the enclosed mass perturbation defined as
\be
m(r) \equiv 4\pi\int_0^r r'^2\d\rho(r')\:dr'.
\ee
If $\d\rho(r)=0$ for $r > R$, the enclosed mass fluctuation $m(r>R)$ is
the total mass fluctuation $\d M$.


Note that
 the $\nabla_i\nabla_j\ph$ term in Eq.~(\ref{eq:phiQS}) is critical
in obtaining this solution since it causes cancelation of the integral terms when integrating by 
parts, leaving
only the boundary terms for the non-linear piece.  
The field solution interior to $r$ has no direct effect on the 
solution at $r$.  Like Newtonian dynamics, only the enclosed mass not the enclosed
field matters.
This property is crucial for maintaining the linearity of the field
solution at large $r$ in the presence of strong non-linearity at small $r$.    
More generally, linearity is a consequence of \refeq{phiQS} satisfying a non-linear Gauss's law and does not require spherical symmetry (see e.g. \cite{Hui:2009kc}).

Since $m \rightarrow \delta M$ at distances beyond which there are no density fluctuations, and assuming that $\ph(\infty)=0$, we can immediately see that the far exterior solution must
be to leading order $\ph \propto 1/r$. Given an increasingly small non-linear term,
this requires
\begin{equation}
\lim_{r\rightarrow\infty} \ph = -{2\over 3\beta}{G \delta M \over r} \,.
\label{eq:linearph}
\end{equation}

For the small $r$ solution,  the key simplification is that the force
modification
$d\ph/dr$ is now an analytic function of the enclosed mass
\begin{equation}
{d \ph \over dr} = {3 \beta r_*(r)\over 4r_c^2}\: g(r/r_*)
\label{eq:forcespherical}
\end{equation}
where the Vainshtein radius of the {\it enclosed} mass is
\begin{equation} 
r_*(r) =\left(\frac{16  G m(r)\: r_c^2}{9\beta^2}\right)^{1/3}
\label{eq:rVgen}
\end{equation}
and
\begin{equation}
g(x) = x [\sqrt{1+x^{-3}}-1] \,.
\end{equation}
Note that in general $r_*$ is a function of $r$ and reflects the \emph{enclosed} mass, and $r_*^3/r^3$ the \emph{average} density,
not the local density.  The latter would be implied by setting
\be
(\nabla_i\nabla_j\ph)^2 = c\:(\nabla^2\ph)^2
\ee 
with $c=$const. in the original field equation. However, doing so would violate the
$r\rightarrow \infty$ limit.    In particular, the far field limit in this approximation would reveal the presence
of a Vainshtein screened mass instead of the true mass perturbation $\delta M$.  
Small scale non-linearity in the density field would then no longer average to give
the required linear perturbations on large scales \cite{Hu:2009ua}.  
Such an approximation or any
that relates the field solution to the local density should not be used in a cosmological
context (cf.~\cite{KW,DGPMpaperII}).

\subsection{Field Profile}
\label{app:prof}

We now assume a top-hat spherical perturbation of radius $R$ with a constant density enhancement $\delta \rho$ [\refeq{drho}]. 
As usual we neglect any compensating underdensity swept out by the prior evolution of
the top-hat.  Since the force modifications within the top-hat only depend on the
enclosed mass $m(r)$ and not the exterior, the compensation does not impact the dynamics.   It can however influence
the field profile $r>R$ and we return to this
point in \S \ref{app:profile}.
For the pure top-hat, $m(r\geq R)=\delta M$ and we can solve for the profile $\ph(r)$ in closed form.

First let us consider the exterior solution at $r >R$.  In the exterior $m$=const. and there
is a single Vainshtein radius $R_* = r_*(R)$.
 Defining a new variable $x = r/R_*$, we can write Eq.~(\ref{eq:forcespherical}) as
\begin{equation}
{d \ph \over d x} = A g(x),
\label{eq:dphidx}
\end{equation}
where 
\begin{equation}
A = {3\beta \over 4} \left( { R_* \over r_c} \right)^2  = {4 \over 3\beta}{ G\dM \over R_*},
\end{equation}
and we can obtain the full exterior field solution as
\begin{eqnarray}
\label{eq:exteriorsoln}
\ph(r) &=& - \int_{r/R_*}^{\infty} {d \ph \over d x} dx\\
&=& {A x^2 \over 2} \left\{ {}_2 F_1[-1/2,-2/3,1/3,-1/x^3] -1 \right\}. \nonumber
\end{eqnarray}
The solution of course recovers Eq.~(\ref{eq:linearph}) in 
the limit of $x \gg 1$
\begin{equation}
\lim_{x \gg 1} \ph = - {A \over 2 x} = - \frac{2}{3\beta}{G\d M \over r} .
\end{equation}
Note that the constant $A$ is $-2 \ph_*$ where $\ph_*$ is the linearized 
field profile evaluated at the Vainshtein radius. 

\begin{figure}[t!]
\centering
\includegraphics[width=0.48\textwidth]{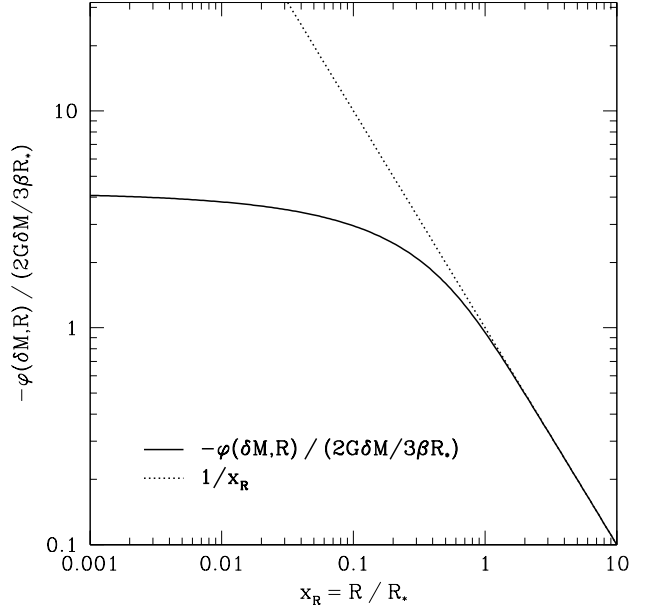}
\caption{Surface field profile $|\ph(\dM,R)|$ (solid curve, see \refeq{phiR}) in units of $(2/3\beta) G\d M/R_*$, 
the linearized value at $R_*$,  as function of $x_R=R/R_*$.  Parameters are
for the sDGP model ($a=1$).
The dotted line shows the linearized solution, $x_R^{-1}$ in these units.
In the Vainshtein limit $x_R\ll 1$, $\ph(\dM,R)$ approaches a constant.
$\ph(\dM,R)$ also determines the binding energy $U_\ph$ (\S~\ref{app:Ebrane}).
\label{fig:phiMR}}
\end{figure}

In the opposite limit $x \ll 1$ 
\begin{equation}
\lim_{x \ll 1} \ph = - A ( C_0-2 x^{1/2}  ).
\label{eq:phiVain}
\end{equation}
The constant piece, 
\begin{eqnarray}
C_{0} &\equiv& A^{-1} \int_0^\infty dx  {d\ph \over dx} \nonumber\\
&=&{\Gamma[1/3]\Gamma[1/6] \over 4\sqrt{\pi}} \approx 2.103 ,
\label{eq:C}
\end{eqnarray}
therefore dominates, and 
\begin{equation}
\lim_{x \ll 1} \ph \approx -{{4 C_{0}\over 3\beta}{ G\dM \over R_*}} = \mbox{const} .
\end{equation}
However
field gradients are determined solely by the $x^{1/2}$ term. 
Since particle dynamics depend on forces and hence field gradients, the 
existence of a constant term in addition to the $x^{1/2}$ term
in the field profile makes an important difference when comparing 
energy conditions like conservation laws and dynamical considerations like
virial equilibrium, as we shall see.
   This distinction is often neglected in the literature
(e.g. \cite{KoyamaSilva,KW}).  
\reffig{phiMR} shows $|\ph(\dM; R)|$ in units of the linearized $(x\gg 1)$ value 
as a function 
of $R/R_*$.  Note the strong suppression of the surface potential and its saturation
for $R \ll R_{*}$.  

\begin{figure}[t!]
\centering
\includegraphics[width=0.48\textwidth]{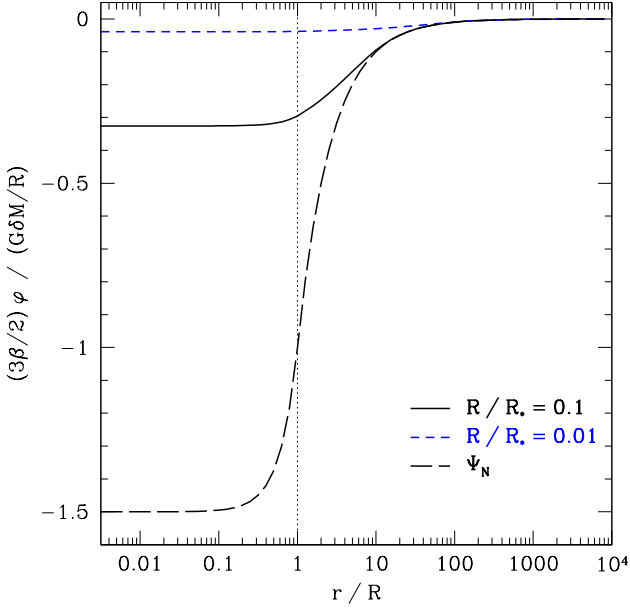}
\caption{Field profile $\ph(r)$ for a top-hat mass of radius $R$ in units
of the Newtonian surface potential $G\d M/R$, for two values of
the Vainshtein radius $R_*/R =10,\,100R$. $\ph$ was scaled by $3\beta/2$
so that the linearized field solution agrees with the Newtonian
potential $\Psi_N$ shown as the long-dashed line.
\label{fig:phiprof}}
\end{figure}

The field interior to the top-hat can be similarly solved.  Since $r_*(r) \propto r$ for 
a constant density profile, Eq.~(\ref{eq:forcespherical}) implies $d\ph /dr \propto r$ and hence
\begin{equation}
\ph(r) - \ph(0)= B r ^2.
\end{equation}
We see from substitution into Eq.~(\ref{eq:sphericalfield}) that  
\begin{equation}
B = {A \over 2 R R_*} g(x_R)  ,
\end{equation}
solves the field equation.
Here
 $x_R = x(R) = R/R_*$.  Note that $d\ph/dr$ is automatically matched at
$r=R$ to the exterior solution Eq.~(\ref{eq:dphidx}).  The remaining condition is that the interior field solution at $r=R$,
\begin{equation}
\ph(R)  = {A \over 2} x_R g(x_R) + \ph(0)
\label{eq:phiR}
\end{equation}
match the exterior solution from Eq.~(\ref{eq:exteriorsoln}).
\reffig{phiprof} shows the resulting $\ph$ profile for different values of $x_R$,
in comparison with the Newtonian potential $\Psi_N$.
Again it is interesting to examine the $x_R \gg 1$ and $x_R \ll 1$ limits
of \refeq{phiR}.
In the former case we regain the linearized (Newtonian) expectation that
the central value is $3/2$ of the surface value
\begin{equation}
\lim_{x_R \gg 1} \ph(0) = -{G \dM \over \beta R} .
\end{equation}
In the opposite, ``Vainshtein'' limit, the central field like the surface field
 is independent of $R$ to leading order
\begin{equation}
\lim_{x_R \ll 1} \ph(0) \approx \ph(R) \approx  -{4 C_{0}\over 3\beta}{G \dM \over R_*} .
\label{eq:vainshteinconst}
\end{equation}
The central field value, which determines the potential
energy associated with the fluctuation is therefore suppressed by $(4C_{0}/3) R/R_*$ from
the linearized expectation. 
Moreover the {\it change} in the field interior or near the object, important for forces, 
is further suppressed by $(R/R_*)^{1/2}$.
We discuss the consequences of these features of the field profile for  virial equilibrium
and the potential
energy in the following section.

The force suppression can be recast in terms of an effective Newton constant
in the Poisson equation for the $\ph$ field.
Note that in the interior solution, the two pieces of the nonlinear term 
in Eq.~(\ref{eq:phiQS}) combine to form 
\begin{equation}
(\nabla^2 \ph)^2 - (\nabla_i \nabla_j \ph)^2  = {2 \over 3} (\nabla^2 \ph)^2 \,.
\end{equation}
Since the field equation is then algebraic in $\nabla^2 \ph$, 
one can solve for $\nabla^2\ph$ to obtain:
\bea
\nabla^2\ph &=& 8\pi \D G_{\rm DGP}(R/R_*)\, \d\rho,\label{eq:delphi2}\\
\D G_{\rm DGP}(x) &=&  \frac{2}{3\beta} g(x) {x^{2}} G \nonumber\\
&=& \frac{2}{3\beta} [\sqrt{1+x^{-3}}-1] {x^{3}} G .
\label{eq:Geff2}
\eea
Hence, in combination with the Newtonian piece, the interior solution can be
phrased as possessing an effective 
$G_{\rm DGP}=G+\Delta G_{\rm DGP}$ modification to gravity.  Furthermore
$x^{-3} \propto \delta$ and so $G_{\rm DGP}$ is a function of the local density inside
the top-hat.
This relation is specific to the interior of a top hat and is {\it not} simply
a spherically symmetric approximation as the exterior solution shows.
More generally, we can see by taking the derivative of Eq.~(\ref{eq:forcespherical}) that
\begin{equation}
\nabla^2 \ph = 8\pi \Delta G_{\rm DGP}(r/r_*)\, \delta\rho(r)  + {2 G m(r) \over r^2 }{d \over d r}[x^2 g(x)] .
\end{equation}
 Note that $\Delta G_{\rm DGP}$, $m(r)$
and $r_*(r)$ are not local functions of the density field but involve the full interior 
profile.
We shall return to this point in \S \ref{app:profile}.

\subsection{Virial Equilibrium}
\label{app:virialeq}

The virial theorem arises from integrating
over space the first moment of the Boltzmann equation, i.e.~from the equation 
of momentum conservation (see e.g. \cite{BinTre87}, \S 4.3).   Despite its usual
association with potential energy, the virial theorem 
is inherently a force balance equation and is the collisionless analogue of
hydrostatic equilibrium.    Thus the virial condition is immune to ambiguities
in the definition of potential energy that we shall discuss in \S \ref{app:Ebrane}.

The virial theorem reads
\be
W \equiv -\int d^3x \:\rho_{m} (\vx)\: \vx\cdot\vn\Psi  = - 2T,
\label{eq:virialtheoremapp}
\ee
where $W$ receives contribution from the Newtonian gravity of the overdensity,
the effective background term, and the brane-bending mode $\ph$. 
For a spherically symmetric top-hat each contribution to $\Psi$ yields
\begin{equation}
W_i =  -3M
\int_0^R \frac{r^2 dr}{R^3}\:r \frac{d\Psi_i}{d r},
\label{eq:W}
\end{equation}
where $\Psi_i$ stands for either $\Psi_N$, $\Psi_{\rm eff}$, or $\Psi_\ph\equiv\ph/2$.
Thus, any constant offsets in $\Psi_i$ do not contribute to $W$.  In particular
the constant term in $\ph$ in the Vainshtein limit of \refeq{vainshteinconst} and its implied potential energy does not
enter into the virial condition.

In the case of the Newtonian contribution, $W_N$ defined in this way is 
given by:
\be
W_N = -\frac{3}{5} \frac{GM\dM}{R},
\label{eq:UN}
\ee
For the effective contribution of the background
$\Psi_{\rm eff}\propto r^2$ and
\be
W_{\rm eff} = - {4\pi G \over 5} [(1+3 w_{\rm eff})\rho_{\rm eff} + \rhob ] M R^2.
\ee
Note that in our convention, we have included the $\rhob$ term in $W_{\rm eff}$
rather than $W_N$. Adding the two contributions yields the familiar
result \cite{HB05}:
\begin{equation}
W_{N}+ W_{\rm eff} = -{3 \over 5} {GM^{2}\over R} - {4\pi G \over 5} (1+3 w_{\rm eff})\rho_{\rm eff} M R^2 .
\end{equation}
Since the trace of the potential energy tensor is defined via forces,
the $\ph$ contribution can be described in terms of $\D G_{\rm DGP}$.
First let us examine the exterior region.
Using \refeq{dphidx}, we obtain
\be
\frac{d\ph}{d r} = \frac{4}{3\beta} \frac{G \dM}{r^2} 
x^2 g(x),
\label{eq:gradphi}
\ee
which using \refeq{Geff} becomes
\be
\frac{d\ph}{dr} = \frac{\D G_{\rm DGP}(r/R_*)\, \dM}{r^2}.
\ee
 In the linear
regime $r \gg R_*$, $\D G_{\rm DGP}$ then reduces to $(3\beta)^{-1}G$.
Note that in the cosmological context, $\D G_{\rm DGP}$ also has a slow
time dependence through $\beta(a)$ [see \refeq{beta}].

Using that in the interior $d\ph/dr \propto r$, we have
\be
W_\ph = -\frac{3 M}{2}
\int_0^R \frac{r^2 dr}{R^3}\:r \frac{d\ph}{d r} = -{3M\over 10}{ d\ph \over d\ln r} \Big|_{r=R} .
\label{eq:Wph}
\ee
Given that $d\ph/dr$ at $r=R$ is determined by the exterior solution, we obtain
\be
W_\ph = -\frac{3}{5} \frac{\D G_{\rm DGP} M\dM}{R},
\label{eq:Wph2}
\ee
where $\D G_{\rm DGP} = G_{\rm DGP}-G$ is the effective gravitational constant
for the force modification [\refeq{Geff}]. Adding all three contributions,
we obtain \refeq{Wtot}.

Specifically in the collapse calculation, we evaluate the virial condition 
\refeq{virialtheoremapp} in terms of the kinetic
and potential energies per unit mass, written in terms of our collapse 
variable $y(\ln a)$.   Defining 
\be
E_{0} = {3 \over 10} M (H_{0}R_{i})^{2}
\ee
where $R_i$ is the initial radius of the perturbation,
we obtain
\bea
\frac{T}{E_{0}} &=& \frac{H^2}{H_0^2} (y' + a/a_i)^2, \nonumber\\
\frac{W_N}{E_{0}} &=& -\Om a^{-3} (\d+1) (y+a/a_i)^2,\nonumber\\
\frac{W_{\ph}}{E_{0}}&=& -\Om a^{-3} \frac{\D G_{\rm DGP}(R/R_*)}{G} \d\:(y+a/a_i)^2,\nonumber\\
\frac{W_{\rm eff}}{E_{0}}&=& -[1+3w_{\rm eff}(a)]\frac{\rho_{\rm eff}(a)}{\rho_{\rm cr,0}} (y + a/a_i)^2.\quad\quad
\eea
We then define the virial radius as the radius during the collapse at which the virial condition is satisfied.  We shall examine approximate techniques for finding this
scale 
through energy conservation in \S \ref{app:Econservation}.  As we shall see, with conventional
definitions of potential energy, the total energy is not strictly conserved, especially during
the initial stages of collapse.

\subsection{Potential Energy Definition}
\label{app:Ebrane}

Let us now consider the potential or binding energy of the top-hat mass.  For the Newtonian contribution, the potential energy is well defined.  By
virtue of the Birkhoff theorem, we can view total mass inside the top-hat as a Newtonian
system in a flat background.  We shall see that neither the potential energy contributed
by the brane bending mode $\ph$ nor the effective forces of the background expansion 
are unambiguous to define as both depend on the exterior cosmological context.  
We therefore follow the convention in the literature in defining them by analogy to
the Newtonian contribution.

The Newtonian calculation of potential energy proceeds by replacing the exterior of the
top-hat with a flat background, the metric analogue to the Newton iron sphere theorem
(see \cite{Pee93} \S 4).   
Removing mass shell by mass
shell from the outside in, we obtain
\be
U =  \int_0^R \: \Psi_{\rm tot}(m(r); r)\:4\pi \rho_{m} r^2 dr
\label{eq:UphiGen}
\ee
where $\Psi_{\rm tot}(m;r)$ denotes the solution for the total gravitational potential for a top-hat with radius
$r$ and mass perturbation $m = (r/R)^3 \dM$ evaluated at $r$.  
This is not to be confused with the total potential at a
radius $r$ interior to the whole top-hat.

Even for the matter contribution there arises an ambiguity in that the contribution of the background matter density across the top-hat is defined only up to 
a constant $\Psi_0$ through $\Psi_{\rm eff}$ of \refeq{Psieff}. 
In other words, the iron sphere theorem applies directly
to forces not potentials and constant offsets do not have any impact on the dynamics.
Therefore 
\be
\Psi_{\rm tot} = \Psi_N + \Psi_\ph + \Psi_{\rm eff} + \Psi_0 \,.
\ee
This constant $\Psi_0$ does not simply introduce a trivial shift in $U$ since even though
it is constant across the top-hat, it is not necessarily constant as we strip away mass shells.
In fact, it is conventional to choose this value to correspond to the result from Newtonian
mechanics
\be 
\Psi_0 = -2\pi  G \bar\rho R^2 \equiv -{3\over 2} {G \overline{M}\over R}
\ee
such that
\be
\Psi_N + \Psi_{\rm eff} + \Psi_0 = -{G M \over R} + {2 \pi G \over 3}(1+3w_{\rm eff})\rho_{\rm eff}R^2 \,.
\ee
The $-GM/R$ piece then corresponds to the Newtonian mechanics result for the 
potential given the total
mass inside the top-hat.   Using the Birkhoff theorem, this is a valid interpretation of 
the cosmological case as well.

 Note that the $\rho_{\rm eff}$ piece cannot properly be considered
a binding energy since its contribution cannot be considered without reference to the
cosmological background.  Even for quintessence models where $\rho_{\rm eff}$ represents
a real energy density, this contribution is supposed
to be smooth within its horizon sized Jeans scale 
regardless of the top-hat collapse and so excising the top-hat and placing the mass in a flat background does not strictly make sense.

Nevertheless, under this convention the binding energy from these three components becomes
\be
U_N + U_{\rm eff} + U_0 = -{3\over 5}{G M^2 \over R} + {2 \pi G \over 5}(1+3w_{\rm eff})M R^2 \,.
\label{eq:UNeff}
\ee
By analogy let us compute the potential energy contribution from $\Psi_{\ph}$,
\be
U_{\ph} = \frac{1}{2} \int_0^R \: \ph(m(r); r)\:4\pi\rho_{m}r^2 dr.
\label{eq:Uphi1}
\ee
Again, $\ph(m; r)$ denotes the {\it exterior} solution of $\ph$  from \refeq{exteriorsoln}
 for a mass $\dM \rightarrow m$ and
radius $R \rightarrow r$.  Note that by making this assumption we are implicitly 
invoking the Birkhoff theorem where it does not in fact strictly apply.   Specifically, 
as mass shells are stripped away, we ignore the impact of the 
{\it underdensity} left behind outside of the body.  As we shall see in \S \ref{app:profile},
this underdensity actually changes the interior profile.    Nonetheless, since the ambiguity
mainly affects the initial stages of collapse when $\delta M \ll M$, it is useful to simply
define \refeq{Uphi1} as the binding energy associated with $\ph$ for energy
 bookkeeping purposes.

Re-expressing in terms of the dimensionless radius $y= r/R$ (not to be confused
with $y$ defined in \refsec{rev} which we shall not use hereafter),
we have $m = y^3\:M$, so that
 $x_{r}= r/r_*(m) = R/R_*= x_R$ is invariant,
while $A({m}) = y^2 A(\dM)$. Therefore,
\be
\ph(m; r) = y^2 \ph(\dM; R),
\label{eq:phiMR}
\ee
where $\ph(\dM; R)$ is the solution for the full mass evaluated at $r=R$
[\refeq{exteriorsoln} at $x=x_R$]. This is the same scaling that one
obtains for a Newtonian potential, $\Psi(\dM; R) = G\dM/R$ and likewise
the potential energy follows the same scaling
\bea
U_{\ph} &=&  \frac{3}{2} M\ph(\dM; R) \int_0^1 y^4 dy \nonumber\\
&=& \frac{3}{10} M\ph(\dM; R).
\label{eq:Uph}
\eea

The important difference for $U_\ph$ is that $\ph(\dM; R)$, the field profile 
at $R$, behaves
very differently in the linearized and Vainshtein limits 
\be
U_\ph = \left \{
\begin{array}{rl}
 -   \frac{1}{3\beta}\frac{3}{5}\frac{G M\dM}{R}, & x_R \gg 1 ,\vspace*{0.2cm}\\
-\frac{2C_{0}}{5\beta} \frac{G M\dM}{R_*}, & x_R \ll 1.
\end{array} \right .
\label{eq:Uphlim2}
\ee
Hence, unlike the Newtonian binding energy, $|U_\ph|$ has a maximum value 
which is reached asymptotically as $x_R=R/R_*$ decreases. Thus, while the Newtonian
binding energy can become an arbitrarily large fraction of the rest
mass energy $\dM$, the energy in $\ph$ is limited to a fraction of
$G\dM/\beta R_* \sim (G\dM/r_c)^{2/3}$ due to the brane-bending mode interactions regardless
of the value of $R$.

\subsection{Potential Energy Usage}
\label{app:PEusage}

Defining a Newtonian based potential energy even though the collapse does not require
a Newtonian interpretation is useful for two interrelated reasons. Firstly it serves
as a bookkeeping device {\it if} the total energy is conserved during the collapse.
Secondly, it can be used to evaluate the virial condition {\it if} it can be simply related to the trace of the potential energy tensor $W$.  We examine to what extent these two
expectations are satisfied given the field $\ph$ and the effect of the background expansion.

Let us define the 
 total energy of the perturbation during collapse as
\be
E = T + U.
\label{eq:Etot}
\ee
Taking the time derivative of \refeq{Etot} and using the
equation of motion of $R(t)$ [Eq.~(\ref{eq:collapse-r})] rewritten using
the top-hat profile as:
\be
\ddot R = -\frac{G_{\rm DGP}\dM}{R^2} - {\FpiG \over 3}[(1+3w_{\rm eff})\rho_{\rm eff}+\rhob]R,
\label{eq:eomR}
\ee
we obtain:
\bea
\frac{dE}{dt} &=& \frac{3}{5}M\dot R\ddot R +  {\partial U \over \partial R} \dot R  + {\partial U \over \partial t}
\\
&=&  {\partial U \over \partial t} , \nonumber
\eea
Here, the partial derivative $\partial U/\partial t$ receives contributions
from evolving quantities in the total potential energy, the sum of \refeq{UNeff} and
(\ref{eq:Uph}). Since $M$ is conserved, there is no violation of energy conservation for
a pure matter system with $\rho_{\rm eff}=0$ and $\ph=0$.  This is a consequence of
adding the $\Psi_0$ offset term to reproduce Newtonian mechanics in the matter terms
including the background.

On the other hand both the $\rho_{\rm eff}$ and $\ph$ contributions have explicit time
evolution across a Hubble time.   Note that violation of energy conservation
due to evolution of $\rho_{\rm eff}$ also
applies to dark energy models where $w_{\rm eff}\ne -1$.  In the $\ph$ term
given by \refeq{Uph}, $\beta$
evolves with the expansion, and as long as $\delta$ is not much greater than 1,
$\delta M$ also evolves during collapse.  More generally these effects occur whenever
the modification to the background (due to modified gravity or the presence
of dark energy) does not match that
of the perturbations. We return to
the impact of energy non-conservation in \S~\ref{app:Econservation}.  Non-conservation
of the Newtonian total energy defined in \refeq{Etot} does not mean a violation in covariant
conservation of energy and momentum.

In order to use the potential energy to assist the evaluation of the virial theorem, we must
relate $U$ to the trace of the potential energy tensor $W$.
It is well known that  for a potential
satisfying $\Psi(r) \propto r^\alpha$ and for which the interior solution is 
$\Psi_{\rm int}(r < R) = \Psi_{\rm ext}(m(r); r)$,
$W_{i} = -\alpha\:U_{i}$. Hence, for potentials satisfying this condition, 
the potential energy determined by the potential itself is of the same order 
as the trace of the potential energy tensor defined by the forces.
This holds for the Newtonian contribution, where we have $U_N=W_N$, and
the effective background contribution, which satisfies $U_{\rm eff}=-W_{\rm eff}/2$.
For the brane-bending mode, the potential is no longer
a pure power law and the distinction 
between $U$ and $W$ leads to interesting consequences in the Vainshtein limit.

\begin{figure}[t!]
\centering
\includegraphics[width=0.48\textwidth]{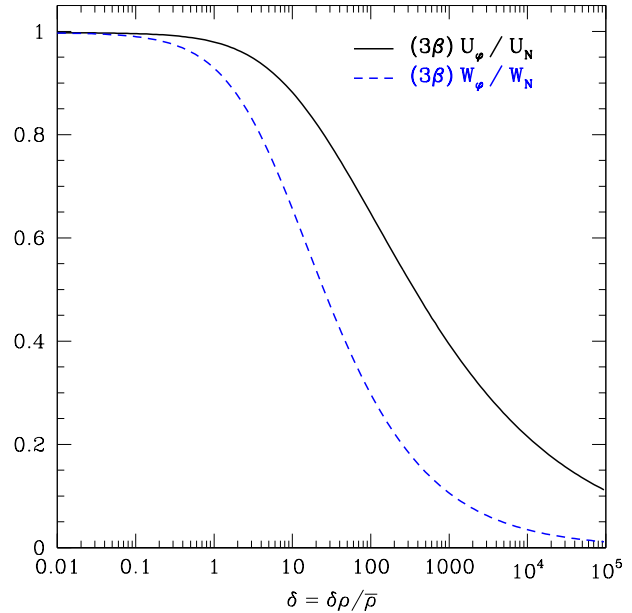}
\caption{Scaled brane-bending mode binding energy $3\beta\times U_\ph$ in units
of the Newtonian binding energy $U_N=-3 GM\dM/5R$ (black solid) as a function of the
overdensity $\d$ for a spherical top-hat mass at $z=0$
in the sDGP cosmology.   The blue
dashed line shows the trace of the potential energy tensor $(3\beta)\times W_\ph$ used in the
virial theorem, \refeq{Wph}, again with respect to the Newtonian value
$W_N = U_N$.
\label{fig:Wbrane}}
\end{figure}

Note that in this $x_R \ll 1$ limit,
the potential energy $U_\ph$ is dominated by the constant term in
\refeq{phiVain}, while the contribution from the $x^{1/2}$ part
of the profile which determines forces is much smaller.
Correspondingly, the assumption that the trace of the potential energy tensor 
$W$ is of order the potential energy $U$ is not valid for the $\ph$ contribution, as the change in the potential across the body is much smaller than the
potential depth itself. The relationship between $W_\ph$ and $U_\ph$ follows
from \refeq{Wph} and~(\ref{eq:Uph}):
\be
W_\ph = - {d \ln \ph \over d\ln r} \Big|_{r=R} U_\ph = -{d \ln U_\ph \over d\ln R} U_\ph .
\ee
In the $R \gg R_*$ limit $\ph \propto r^{-1}$ and $W_\ph = U_\ph$ as usual
but in the $R \ll R_*$ limit, $\ph \approx$ const. and the trace of the potential tensor
is highly suppressed compared to the potential energy.
\reffig{Wbrane} shows $U_\ph$ and $W_\ph$ as a function of the overdensity
$\d=\d\rho/\rhob$ of the perturbation.   Note that if we were to interpret the Vainshtein
effect as simply a modification of $G$ we would infer the wrong energy condition 
at virialization.   We discuss this issue in the next section.

\begin{figure}[ht]
\centering
\includegraphics[width=0.48\textwidth]{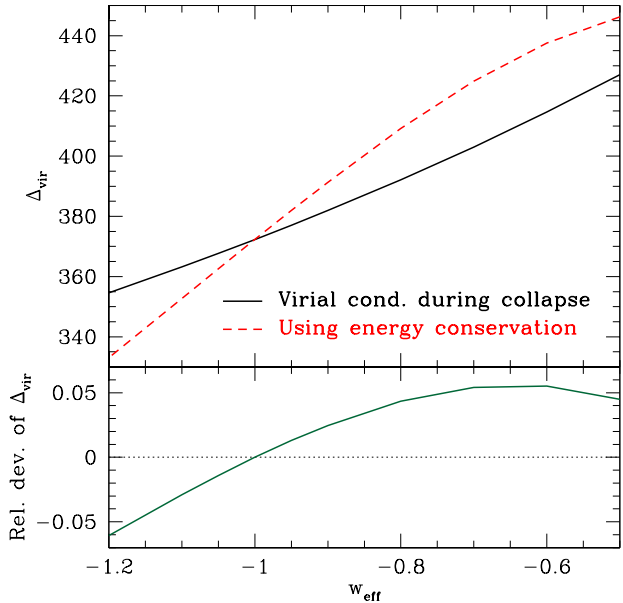}
\caption{Misestimation of the virial overdensity $\Dvir$ in quintessence dark energy models.
$\Dvir$ as defined in \refsec{virial} for collapse
at $a_0=1$, as a function
of the dark energy equation of state parameter $w_{\rm eff}=$const. and 
$\Omega_{\rm eff}= 0.741$ i.e. for standard
gravitational forces (top panel).  The solid line shows $\Dvir$ determined
by evaluating the virial condition during collapse,
the approach adopted here, while the dashed line shows the usual calculation
using standard energy conservation as described in \refapp{Econservation}.  Both
calculations agree for $w=-1$, i.e. $\rho_{\rm eff}=$const.  The lower panel
shows the relative deviation between the two, $(\Dvir^{\rm std}-\Dvir)/\Dvir$. 
\label{fig:Dvir-w}}
\end{figure}
\subsection{Misestimating Virial Overdensity}
\label{app:Econservation}

Most commonly, the virial condition \refeq{virialtheorem} is evaluated using energy
conservation. 
At virialization,
\begin{equation}
T(\Rvir) = -\frac{1}{2}W(\Rvir)
\end{equation}
and the total energy $E = T(\Rvir) + U(\Rvir) = U(\Rvir) - W(\Rvir)/2$. 
Since at turn-around ($R=\Rta$) the kinetic energy vanishes,
we have $E = U(\Rta)$. Assuming energy conservation we obtain
\be
U(\Rta) - U(\Rvir) + \frac{1}{2}W(\Rvir) = 0.
\label{eq:virialE}
\ee
By further assuming a relationship between the potential energy $U$ and the
trace of the potential tensor $W$ we can solve for $\Rvir/\Rta$.
There are therefore two ways by which this association can go wrong: if $E$ is not
conserved and if an incorrect relation between $U$ and $W$ is employed.

Let us begin by examining the first issue.
Energy is not strictly conserved in any model where either $w_{\rm eff} \neq -1$
(including quintessence),  modifications to gravity are 
time variable, 
or force modifications are only generated by the perturbed mass $\delta M$.  

Evaluating \refeq{Etot} during collapse, we found that
in the effective dark energy model QCDM (which has the same $H(z)$ as sDGP),
energy conservation is violated by $\sim 3$\% from turn-around to collapse.
While the violation of energy conservation in dark energy models thus seems
to be minor, it does influence the virial overdensity due to the sensitivity
of $\Dvir$ to $\Rvir$ and $\avir$. \reffig{Dvir-w} shows $\Dvir$
as function of a constant dark energy equation of state $w_{\rm eff}$ and 
$\Omega_{\rm eff}=0.741$, determined by evaluating the virial condition during collapse
(our approach) and using energy conservation (e.g., \cite{HB05}). 
While both approaches agree
for $w_{\rm eff}=-1$ as expected, there are clear differences as soon as $w_{\rm eff}\neq -1$.
Note that these differences are of the same order as the difference
$\Dvir(w_{\rm eff}) - \Dvir(w_{\rm eff}=-1)$.

Our approach does not rely on exact energy conservation and one might
thus expect the $\Dvir$ obtained in this way would lead to a better match
to observables.    Since quantities such as the mass function are typically simply fit to simulations with a given definition of overdensity, this is in part an issue of semantics.
However use of a more physically motivated scaling might lead to 
a more universal form for the mass function or one that scales more simply with parameters
of the theory.  It would however be necessary to
compare with N-body simulations of $w_{\rm eff}\neq -1$ dark energy models and that is
beyond the scope of this work.  
 Here we simply note that  the dependence on $w_{\rm eff}$ of $\Dvir$
determined using our approach is smaller than that of the usual definition and so it would
predict a more universal scaling with a fixed overdensity than the standard approach.

The deviations from $E=\rm const.$ are much larger in the DGP modified force
case, where evolving forces and $\dM$ lead to stronger evolution of $E$.
Here, differences in the total energy between turn-around and collapse
are $10-15$\%.  Thus, the choice of procedure for determining $\Dvir$
becomes even more important.  Again, we found that our approach (greatly) reduces
the dependence of $\Dvir$ on the evolution of the modified forces.

Finally, we consider the effect of assuming an incorrect relation between
$U$ and $W$ in \refeq{virialE}.  For example, it is tempting to just
set $U_\ph = W_\ph = \D G_{\rm DGP}/G\times W_N$, i.e., ignoring the
$R$-independent term in $U_\ph$.  However, since $\D G_{\rm DGP}\rightarrow 0$ 
in the late stages of collapse ($\d\rightarrow \infty$), $U_\ph$ would
erroneously be set to $0$ in this approximation.  For $\beta$ of order
unity, this leads to apparent violations of energy conservation at the
level of 30\%.  When using energy conservation in the presence of the
$\ph$ field, it is thus crucial to take into account the differences
between $U_\ph$ and $W_\ph$.

\begin{figure}[t!]
\centering
\includegraphics[width=0.48\textwidth]{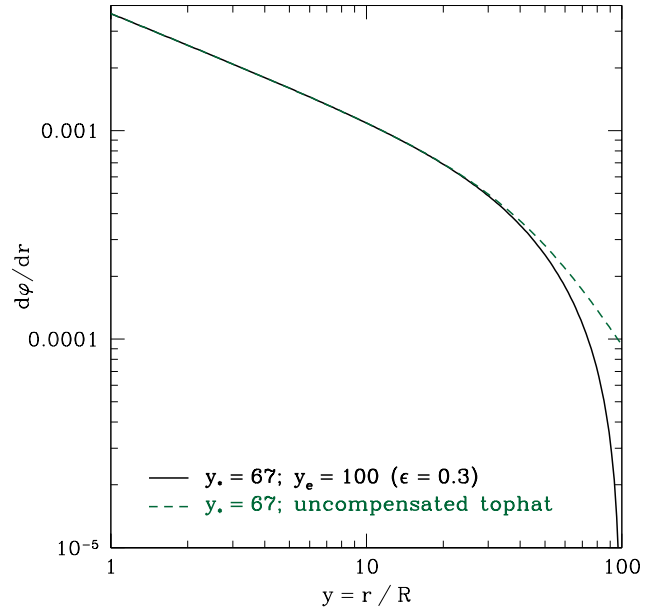}
\caption{Field gradient $d\ph/dr$ for a compensated top-hat profile (solid) in units of
$2/(3\beta) G\d M/R^2$ as a function of 
$y=r/R$. $y_e=R_e/R$ is set to 100 (corresponding to $\d = 10^6$), and 
$y_* = R_*/R=67$, corresponding to $\epsilon=0.3$ valid for the sDGP model.
The dashed line shows $d\ph/dr$ for the same top-hat mass but with 
uncompensated profile.  Near $R$, force modifications from $\ph$ are not affected by
compensation.
\label{fig:dphidrCTH}}
\end{figure}

\subsection{Compensated Top-hat Profile}
\label{app:profile}

Finally we study one example of a density profile beyond the pure top-hat,
the \emph{compensated top-hat} profile.
In the cosmological context, a collapsing top-hat perturbation
sweeps out ``empty space'' and
in fact has the following density profile:
\be
\rho(r) - \rhob = \left\{
\begin{array}{rl}
\rhob \d, & r \leq R,\\
-\rhob, & R < r \leq R_e,\\
0, & r > R_e,
\end{array} \right .
\label{eq:drhotrue}
\ee
where $R_e = R_i a/a_i$ is the physical radius today corresponding to the radius
of the perturbation at an early time $a_i$ when $\d \ll 1$.
We continue to call the total mass and mass perturbation enclosed at $R$ as $M$ and
 $\delta M \equiv m(R)$ respectively.

In terms of the scaled radial coordinate $y\equiv r/R$, the full description for
the enclosed mass perturbation becomes
\be
m(r) = \left\{
\begin{array}{rl}
\d M y^3, & y \leq 1,\\
M [ 1-(y/y_e)^3],   & 1 < y \leq y_e,\\
0, & y> y_e.
\end{array} \right .
\label{eq:mr}
\ee
Importantly, forces at a given radius only depend on the enclosed mass $m(r)$  through  \refeq{forcespherical}.
Given that for $r<R$ the enclosed mass of the compensated and uncompensated
top-hat are the same, compensation has no impact on the interior dynamics of 
collapse.  Likewise, from the definition of the potential energy
tensor \refeq{W}, we see then that $W_\ph$ is unchanged from that of
the pure top-hat profile and hence the virial condition is unmodified. 

  Naively, one might assume that as long as $\d \gg 1$, the
compensation has little effect on the binding energy or gravitational potential $\Psi$ in the interior as well but we shall see that
this is not necessarily so for the brane-bending contribution  due to the Vainshtein
suppression.

In the exterior ($y>1$) forces from $\ph$ are modified by the compensation as
\bea
\label{eq:dphidrCTH}
\frac{d\ph}{dy}&=& R\:\frac{d\ph}{dr}
\\
& = & \left (\frac{2}{3\beta}\frac{G\d M}{R}\right ) \frac{2y}{y_*^3}\left [\sqrt{\left(\frac{y_*}{y}\right)^3\frac{y_e^3-y^3}{y_e^3-1}+1}-1\right ]. \nonumber
\eea
In the limit $y_e \gg y \ge 1$, the forces are the same as in the uncompensated profile
(see Fig.~\ref{fig:dphidrCTH}).
Since $y_e^3 = 1+\delta$, forces are unchanged near the body, $y \sim 1$,
as long as $\delta \gg 1$.  
Even for $\delta < 1$, this modification does not introduce any physical effect on the collapse
 since there is no mass in the exterior region $1 < y < y_e$ that
could be moved by the modified forces.   Hence an initial top-hat profile will
remain a top-hat during collapse.

Now let us look at the effect of compensation near the Vainshtein scale of the mass, 
$y_* \equiv R_*/R$. Note that 
\begin{equation}
{y_*^3 \over y_e^3} = \epsilon {\delta \over 1+\delta}
\end{equation}
where $\epsilon^{-1}$ is the density threshold beyond which the Vainshtein mechanism
operates as defined in \refeq{eps}.  Unless this density threshold also satisfies
 $\epsilon^{-1} \gg 1$, compensation effects will change how the profile saturates,
since $y_*$ will be comparable to $y_e$.
Correspondingly in \refeq{dphidrCTH}, both $\delta \gg 1$ and $\epsilon \ll 1$ are
necessary for $d\ph/dy$ to recover the uncompensated result at the Vainshtein
scale $y_*$. Given that the top-hat $\ph$ profile within $R_*$ is controlled 
by its value at $R_*$, we expect the $\ph$ profile itself to be modified
near the body by the density compensation unless
$\epsilon \ll 1$.

\begin{figure}[t!]
\centering
\includegraphics[width=0.48\textwidth]{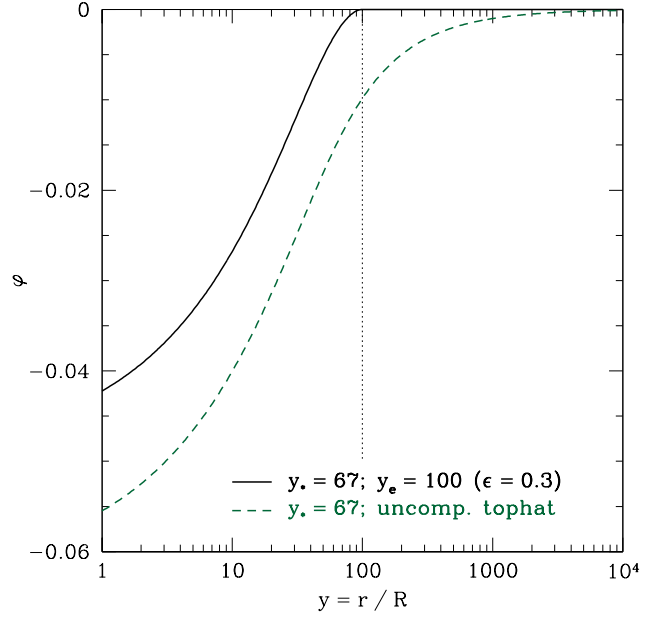}
\caption{The exterior $\ph$ profile in units of $2/(3\beta) G\d M/R$ as 
function of the scaled radius $y$ for a compensated top-hat profile (solid).
$y_e=100$ and $y_*=67$ as in \reffig{dphidrCTH}.
The field goes to 0 at $y=y_e$ (dotted vertical line). The dashed line
shows the profile for an uncompensated top-hat with the same mass and radius.
Note that the $\ph$ field for the two profiles differs at all radii, reflecting
the fact that $\ph(r)$ is not simply determined by the enclosed mass at $r$.
\label{fig:phiCTH}}
\end{figure}

More specifically, let us consider the linearized ($y_* \ll y$) and  
Vainshtein ($y_* \gg y$)
limits as before.  First note that even the Newtonian
force $d\Psi_N/dr$ for the compensated top-hat at $y>1$ is modified as
\be
R\:\frac{d\Psi_N}{dr} = \frac{G\:m(r)}{R} y^{-2} = \frac{G\d M}{R}\frac{y_e^3-y^3}{y_e^3-1} y^{-2}.
\label{eq:dPsidrCTH}
\ee
In the linearized limit, \refeq{dphidrCTH} reduces
to 
\begin{equation}
\lim_{y\gg y_*} {d\ph\over dr} = {2 \over 3\beta} {d\Psi_N \over dr}
\end{equation}
 as expected. 
In the Vainshtein limit of $y_* \gg y (>1)$, and in which case $y\ll y_e$ as well, the force contributions take the form
\be
\lim_{y \ll y_*} \frac{d\ph}{dy} = \left (\frac{2}{3\beta}\frac{G\d M}{R}\right )
\frac{2}{\sqrt{y}} y_*^{-3/2} \left(\frac{y_e^3-y^3}{y_e^3-1}\right)^{1/2}.
\ee
We thus recover the leading $r^{-1/2}$ behavior of the force in this limit.
 \reffig{dphidrCTH} shows
$d\ph/dy$ as function of $y$ for different values of $y_*=R_*/R$.
Note that unlike the
Newtonian case, the force given by $d\ph/dy$ differs from that of an 
equivalent top-hat with radius $r$ and enclosed mass fluctuation $m(r)$
for any $r>R$. This behavior cannot be described by a simple $G_{\rm DGP}(\d)$ 
parametrization.

Next as with the pure top-hat, we can solve for the whole field profile 
given the boundary condition $\ph(R_e)=0$ by integrating
\be
\ph(r>R) = -\int_{r/R}^{y_e} \frac{d\ph}{dy} dy.
\ee
We then obtain
\bea
 \ph(y) = -\left (\frac{2}{3\beta}\frac{G\d M}{R}\right )
{y_e^2 \over y_*^{3}} 
  \left [ {y^2 \over y_e^2} -1 + 
 F(1,\epsilon)
- F({y \over y_e},\epsilon)
\right ] ,\nonumber
\label{eq:phiCTH}
\eea
where
\begin{equation}
F(x,\epsilon) 
\equiv 4 \sqrt{\epsilon x}\, {}_2 F_1[-1/2,1/6;\,7/6;\,x^3(1-\epsilon^{-1})] .
 \end{equation}
Note that this is a slightly different hypergeometric function than that in the
pure top-hat field solution. 
 \reffig{phiCTH} shows $\ph(y)$ vs. $y$ for a fixed
overdensity $\d = 10^6$ for two different values of $y_*$ (or, equivalently,
$\epsilon$). 
Again, we obtain the expected scaling in the
limiting cases. For the linear limit $y \gg y_* $, we have
\bea
\lim_{y \gg y_*} \ph(y) &{=}& \left (\frac{2}{3\beta}\frac{G\d M}{R}\right )
\left [ -\frac{y^2}{2\d}-\frac{\d+1}{y \d} + \frac{3}{2}\frac{(\d+1)^{2/3}}{\d}\right ]\nonumber\\
&=& \frac{2}{3\beta}\Psi_N,
\eea
proportional to the Newtonian potential for a compensated top-hat with
the same boundary condition, $\Psi_N(R_e)=0$. As expected, for $\delta \gg 1$ the profile matches the uncompensated case until 
$y$ approaches $y_e$.

\begin{figure}[t!]
\centering
\includegraphics[width=0.48\textwidth]{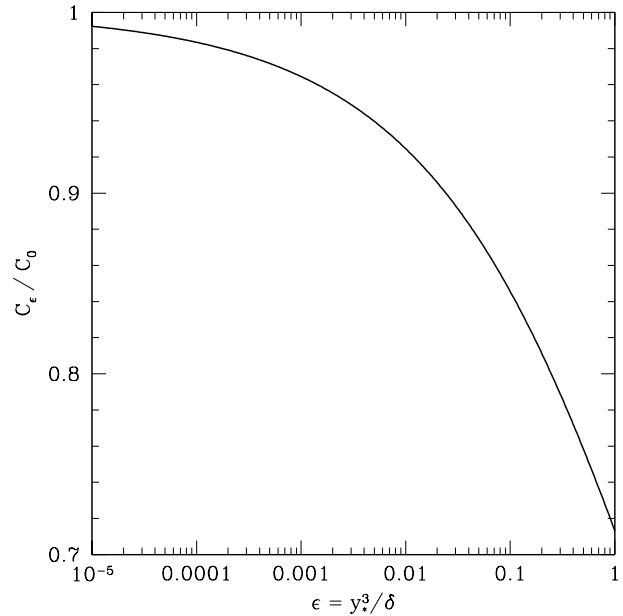}
\caption{Reduction in the surface potential
due to compensation in the Vainshtein limit, $C_{\epsilon}/C_{0}$,
as a function of $\epsilon = y_*^3/\d$. In the limit $\epsilon \ll 1$, $y_* \ll y_e$
and the compensation has little effect on the $\ph$ profile.  This quantity also
controls the reduction in binding energy. \label{fig:phiInf}}
\end{figure}

 In the Vainshtein limit
$y_*\approx \epsilon^{1/3} y_e \gg 1$ and the profile becomes
\be
\lim_{y \ll y_*} \ph(y) {=} - \left (\frac{4}{3\beta}\frac{G\d M}{R_*}\right )
\left[ C_{\epsilon} - 2  \sqrt{y/y_*}\right] 
\label{eq:phiCTHV}
\ee
where
\be
C_{\epsilon} = {F(1,\epsilon)-1 \over 2 \epsilon^{2/3}}.
\ee
In the limit $\epsilon \rightarrow 0$, $C_{\epsilon}=C_{0}$, and 
the profile returns to the uncompensated form of \refeq{vainshteinconst}.
In
\reffig{phiInf} we show $C_{\epsilon}/C_{0}$ the reduction in the surface potential $\ph(R)$ 
due to the compensation in the Vainshtein limit.

One can also define an alternate definition of the binding energy of the perturbation 
$\delta M$.   Suppose that we again define the binding energy $U_\ph$ as in \refeq{Uphi1} by
removing shell by shell of the mass to $R_{e}$.  However, in this case we properly
account for the impact of the exterior.   As each shell is removed to $R_e$, it fills
in the mass deficit so that $R_e$ decreases  in such a way as to keep
$y_{e}=R_{e}/R$ constant.

Thus, the only part of \refeq{phiCTH} which scales nontrivially is the
prefactor $\d M/R \propto R^2$ as in the case of the pure top-hat. 
\refeq{Uph} strictly holds for the compensated top-hat 
\be
 U_i = \frac{3}{10}M \Psi_i(\d M, R)
 \label{eq:Uicompensated}
\ee
for the Newtonian and $\ph$ contributions.
In particular, in the Vainshtein regime, the modification in $U_{\ph}$ from 
the pure top-hat is given by the constant piece in the profile which scales
with $\epsilon$ as in Fig.~\ref{fig:phiInf}.

In fact, this derivation unlike that for the pure top-hat is fully self-consistent in that the
form of the profile, including the background contribution, is self-similar as the
shells are removed.  Furthermore the end result is a compensated top-hat of
$R \rightarrow 0$, i.e. an unperturbed universe with no source to $\ph$.
In this view the binding energy associated with $\delta M$ is the energy required
to eliminate the mass perturbation rather than the mass.

Unfortunately, this definition still does not fully resolve the ambiguities associated
with the potential energy definition.  We still need to account for the potential energy due to the
background density.  In particular, in this definition the Newtonian binding energy also only accounts for the perturbation and hence
using it in the definition of total energy would not obey strict energy conservation during
the collapse, especially for $\delta < 1$.  
We have seen that 
in the literature the definition of potential energy is 
mainly used in conjunction with energy conservation to simplify the calculation of the virial radius. 
In this context, the original definition of binding energy as that of a pure top-hat with no contribution from the exterior is more useful. 
In this case it is important to keep in mind however that 
modified forces as well as evolving dark energy density generally imply
a violation of energy conservation (\refapp{PEusage}).

\bibliographystyle{arxiv_physrev}
\bibliography{DGPM}

\end{document}